\documentclass{aa}
\usepackage{graphics}
\usepackage{psfig}
\begin{document}

   \thesaurus{ 11.16.1;    
               11.06.2;    
               04.19.1;    
               11.19.2;    
	       11.19.3;    
               03.13.2}    
   \authorrunning{P. G. P\'erez-Gonz\'alez et al.} \titlerunning{Optical surface photometry of the UCM Survey}
   \title{Optical photometry of the UCM Lists I and II}

   \subtitle{II. $B$ band surface photometry and morphological discussion}

   \author{P. G. P\'erez-Gonz\'alez$^1$, J. Gallego$^2$, J. Zamorano$^3$ \and A. Gil de Paz$^4$ }

   \offprints{P.G. P\'erez-Gonz\'alez}

   \institute{Departamento de Astrof\'{\i}sica, Universidad
   Complutense de Madrid, Av. Complutense s/n. E-28040 Madrid Spain\\
   e-mails: pag@astrax.fis.ucm.es,jgm@astrax.fis.ucm.es,jaz@astrax.fis.ucm.es,gil@astrax.fis.ucm.es }

   \date{Received August 2 2000. Accepted October 20 2000.}

   \maketitle

   \begin{abstract} 

	We present Johnson $B$ surface photometry for the UCM Survey
	galaxies. One-dimensional bulge-disk decomposition is
	attempted, discussing on fitting functions and computational
	procedures. The results from this decomposition, jointly with
	concentration indices and an asymmetry coefficient, are
	employed to study the morphological properties of these
	galaxies. We also compare our results with the previous
	morphological classification established using Gunn $r$
	imaging data and with other samples of galaxies. No major
	statistical differences in morphology are found between red
	and blue data, although some characteristics such as size and
	luminosity concentration vary. We find a correlation between
	luminosity and size. Several parameters are used to segregate
	the objects according to their morphological type.

      \end{abstract}

%

\section{Introduction}

        The Universidad Complutense de Madrid Survey (UCM Survey List
        I, Zamorano et al. \cite{Zam94}; List II, Zamorano et
        al. \cite{Zam96}; List III, Alonso et al. \cite{Gar99})
        constitutes a representative and fairly complete sample of
        current star-forming galaxies in the Local Universe (Gallego
        \cite{Gal99}). Its main purposes are to identify and study new
        young, low metallicity galaxies and to quantify the properties
        of the current star formation in the Local Universe. Another
        key goal is also to provide a reference sample for the studies
        of high-redshift populations, mainly dominated by star-forming
        galaxies (see, e.g., Hu et al. \cite{Hu98} and Steidel et
        al. \cite{Ste99}).

	Specific details of the UCM sample concerning spectroscopic
	properties were presented in Gallego et al. (\cite{Gal95},
	\cite{Gal96}, \cite{Gal97}). Photometric properties in the
	Gunn $r$ band can be found in Vitores et al. (\cite{Alv96a},
	\cite{Alv96b}) and near-infrarred data are available in
	Alonso-Herrero et al. (\cite{Alm96}) and Gil de Paz et
	al. (\cite{Gil99}).

	In P\'erez-Gonz\'alez et al. (\cite{yor}, hereafter Paper I),
	integrated Johnson $B$ photometry for the whole sample was
	presented. In this paper we will study the spatial
	concentration of the $B$ luminosity in these objects. This
	will allow us to obtain information about morphology and the
	different structures embedded in the galaxies. The effects of
	the photometric band on the determination of galaxy morphology
	will be evaluated through the comparison of the results
	presented here with those achieved with the Gunn $r$ data
	(Vitores et al. \cite{Alv96a}, \cite{Alv96b}).

	The study of large-scale characteristics of the galaxies
	involves the quantitative measurement of structural parameters
	and light distribution. These parameters, which must describe
	the different components of galaxies (i.e., bulges, disks,
	bars), may be used to perform a morphological classification
	of the objects.

	In this sense, we will attempt bulge-disk decomposition in one
	dimensional surface photometry radial profiles. Calculation of
	concentration indices and an asymmetry coefficient will also
	be done. All these data will be utilized to classify the
	galaxies according to their Hubble type.

        The paper is structured as follows: we introduce the sample of
        galaxies and the Johnson $B$ observations in
        Sect.~\ref{obs}. In Sect.~\ref{data} the method used in this
        surface photometry study is detailed, including the
        explanation of the procedures followed in the bulge-disk
        decomposition and in the calculation of concentration indices
        and the asymmetry coefficient. The results and discussion
        about morphology are located in Sect.~\ref{results}. Finally,
        Sect.~\ref{correlations} presents the correlations found
        between several photometric parameters. A Hubble constant
        $H_{0}$=50\,km\,s$^{-1}$\,Mpc$^{-1}$ and a deceleration
        parameter $q_0$=0.5 have been used throughout this paper.

\section{The sample}
\label{obs}
	
	The present work refers to the 191 galaxies within the lists I
	and II of the UCM Survey. The main features of each galaxy as
	well as the observation parameters and reduction techniques
	were listed in Paper I.

	The UCM Survey galaxies lie at an averaged redshift of
	0.026. The sample is dominated by low-excitation,
	high-metallicity starburst-like galaxies (57\% of the sample);
	high-excitation, low-metallicity HII-like galaxies are also
	present (32\%) with a fraction of AGN objects (8\%) (Gallego
	et al. \cite{Gal96}). The averaged $B$ magnitude and standard
	deviation of the sample are 16.1$\pm$1.1 and the mean $B-r$
	colour 0.73$\pm$0.41 (P\'erez-Gonz\'alez et al. \cite{yor}),
	corresponding to a morphological type of Sbc, according to
	Fukugita et al. (\cite{Fuk95}).

\section{Data analysis}
\label{data}

\subsection{Surface photometry}
\label{data.1}
	
	Surface photometry was performed with the {\it ellipse} task
	within the IRAF\footnote{IRAF is distributed by the National
	Optical Astronomy Observatories, which is operated by the
	Association of Universities for Research in Astronomy,
	Inc. (AURA) under cooperative agreement with the National
	Science Foundation.} {\sc stsdas}.{\sc isophote} package.
	
	The {\it ellipse} task matches elliptical isophotes to galaxy
	light intensity.  The technique employed is described in
	Jedrzejewski (\cite{Jed87}). It consists on an iterative
	method that fits ellipses to the intensity of the images at a
	given semi-major axis. Once the isophote is fitted, the
	semi-major axis is changed (increased or decreased). This task
	was run interactively always starting from the same semi-major
	axis (the one corresponding to 3 arcsec to avoid the part of
	the galaxy profile dominated by seeing). First, we proceeded
	outwards until we reached twice the Kron radius (Kron
	\cite{Kron80}) and then inwards.
	
	Intensities were converted into surface brightnesses and
	plotted against the semi-major axes of the isophotes.  No
	correction for inclination ($i$) was attempted before fitting
	the profiles due to the great variety of uncertainties that
	the determination of $i$ involves. These surface brightness
	profiles, together with the bulge-disk adjustments explained
	in the Sect.~\ref{meth}, are available via anonymous ftp at
	the site 147.96.22.14.

	Along with the surface brightness profiles, the method
	mentioned above also provides the ellipticity $\epsilon$ and
	position angle PA of each isophote. For each image, mean
	$\epsilon$ and PA were calculated with the values of the
	isophotes between 23 and 24 mag$\cdot$arcsec$^{-2}$ and are
	listed in Table \ref{fe}. Since the outer isophotes of many of
	our galaxies were distorted by different structures, such as
	bars, rings, spiral arms, bright HII regions, some of these
	averages were corrupted, so a visual inspection of each image
	was carried out in order to exclude from the averaging the
	distorted zones and get more indicative values.

\subsection{Bulge-disk decomposition. The method}
\label{meth}
	
	Traditionally, galaxy light distributions have been studied
	through the decomposition in distinct components (Kent
	\cite{Ke85}, de Jong \cite{Jong96b}, Vitores et
	al. \cite{Alv96a}, Baggett et al. \cite{Bag98}, among
	others). Most methods are based on the assumption of specific
	functions. Ideally, these functions should have a physical
	background, being connected with the formation and evolution
	of galaxies. Unfortunately this is a very hard task so authors
	commonly use empirically derived functions.

	Light distributions of spiral galaxies are commonly modeled
	using two components: a central concentration of luminosity
	(the bulge) and an outer plane structure (the disk). This
	simple scheme can be far from the real component mixture of
	the galaxy. Features such as bars, rings or bright starbursts
	affect dramatically the light distribution and make bulge-disk
	decomposition a nearly impossible task. These features are
	supposed to be more frequent in late Hubble type galaxies and
	extremely relevant in starburst galaxies, becoming dominant at
	high-redshifts.

	Bulge-disk decomposition can be undertaken using several
	techniques and fitting functions. Several authors are now
	using the entire galaxy image to perform two dimensional
	fittings of the flux (see, e.g., de Jong \cite{Jong96b}); this
	technique is better for galaxies with peculiar structures such
	as bars or rings, which are masqueraded in the azimuthally
	averaged plots.

	We have carried out the morphology study of the sample using
	one-dimensional surface brightness profiles. These profiles
	were checked visually in order to exclude from the fitting
	algorithm those regions dominated by artifacts, which are
	revealed through bumps and dips in the radial
	profiles. Besides, the algorithm only utilizes the points with
	$\mu$ lower than the detection threshold, which was measured
	as the surface brightness corresponding to the standard
	deviation of the sky; the values of this threshold ranged from
	24 to 26 mag$\cdot$arcsec$^{-2}$, depending on the observation
	campaign. Some of the galaxies showed very irregular
	morphologies and extremely perturbed profiles due to
	interaction companions or starbursts; consequently, these
	galaxies were excluded from this bulge-disk study.

	A great variety of fitting functions are available in the
	literature. Some authors adjust exponential laws to both bulge
	and disk or other more complicated functions. We have
	attempted the decompositions using the empirical bulge law
	established by de Vaucouleurs (\cite{Vau48}):

	\begin{equation}
		\mu=\mu_e+8.33\cdot \left( \left( \frac{r}{r_e} \right)^{1/4} - 1 \right)
	\end{equation}

	\noindent and the classical exponential law for the disk
	(Freeman \cite{Fre70}):

	\begin{equation}
		\mu=\mu_0+1.09\cdot \left( \frac{r}{d_L} \right)
	\end{equation}

	\noindent where $\mu$ stands for the surface brightness, $r$
	for the radius, $r_e$ for the bulge effective radius (that
	containing inside half of the total light of the bulge
	component), $\mu_e$ the bulge effective surface brightness,
	$d_L$ and $\mu_0$ the disk scale length and central surface
	brightness.
	
	The choice of the classical $r^{1/4}$ and exponential fitting
	functions allows us to compare our results with the Gunn $r$
	study and with most of the data found in the literature.

	During the performance of bulge-disk decomposition, special
	care should be taken when dealing with the inner parts of the
	galaxy profile, since these zones are affected by atmospheric
	seeing. Most authors exclude from the fit the part of the
	galaxy dominated by seeing (e.g., Baggett et al. \cite{Bag98},
	Schombert \& Bothun \cite{Sch87}, Chatzichristou
	\cite{Cha99}). To account for this effect, we used in the
	fitting procedure a seeing-convolved formula for the light
	profile in the inner parts of the galaxy (Pritchet \& Kline
	\cite{PriKl81}). This procedure copes with the uncomfortable
	$r^{1/4}$ bulge law, that tends to infinity as $r$ approaches
	0. Assuming radial symmetry and a gaussian description of the
	PSF, the seeing convolved profile can be expressed as:

	\begin{equation}\label{seeing}	
	I_c(r)=\sigma^{-2}e^{-r^2/2\sigma^2}\int_{0}^{\infty}I(x)\,I_0(xr/\sigma^2)\,e^{-x^2/2\sigma^2}x\,dx
	\end{equation}
	
	\noindent where $I_c(r)$ is the seeing-convolved intensity,
	$\sigma$ the dispersion of the seeing gaussian PSF, $I(x)$ the
	sum of the bulge and disk intensities and $I_0$ the zero-order
	modified Bessel function of the first kind. Seeing dispersions
	were measured on several field stars for each image; the
	averaged seeing value was $1\farcs5\pm0\farcs4$, ranging from
	$0\farcs9$ to $2\farcs0$.

	The main problem involved with seeing is the determination of
	the seeing-dominated zone of the profile, where Eq.
	(\ref{seeing}) has to be used. This parameter was set free
	until a best fit was achieved.

	The decomposition procedure followed to obtain the bulge and
	disk parameters is the following:

	\begin{itemize} \item All the fitting subroutines need a first
	guess for the bulge and disk parameters $\mu_e$, $r_e$,
	$\mu_0$ and $d_L$. To calculate them we separated the bulge
	and disk dominated regions; the first one should be located in
	the centroid of the galaxy (with the most central zones
	dominated by seeing) and should be linear in a surface
	brightness versus $r^{1/4}$ plot; the zone dominated by the
	disk must be in the outer parts of the galaxy and should be
	linear in a $\mu$ versus $r$ plot. This decomposition provided
	a first estimate for the bulge and disk parameters.\item The
	next step is to fit the two components simultaneously. This
	was done using a $\chi^2$ minimization performed with the
	simplex method. Before this minimization, we proved several
	solutions in the parameter space around the data acquired in
	the first step in order to avoid local minima of the $\chi^2$
	function. The simplex algorithm needs five sets of initial
	parameters that were built with the best of the previous
	values, varying them randomly.\item One final step was
	performed in order to calculate the best set of fitting
	parameters and their corresponding errors. Each data value in
	the surface brightness profile was varied randomly according
	to a gaussian distribution; the sigma of this gaussian was the
	standard deviation of the point calculated with the {\it
	ellipse} task. Each new profile was refitted using the simplex
	method. This process was repeated 1000 times. Then those fits
	with values of the function $\chi^2$ between the lowest value
	$\chi^2_{min}$ and 3 times $\chi^2_{min}$, were selected from
	the 1000 iterations. The final set of parameters were the
	averaged values of the latest and the errors corresponded to
	the standard deviations of these fits.\end{itemize}

	Equal weights were used for all the points during the
	fits. The outermost points of the profiles have larger errors
	due to the uncertainties in the determination of the sky,
	artifacts, etc. This should lead to assign greater weights to
	the innermost points, as some authors do in the literature
	(Baggett et al. \cite{Bag98}, Chatzichristou \cite{Cha99},
	Hunt et al. \cite{Hun99b}). However, in our profiles there are
	more points in the inner parts of the plots than in the outer
	ones; when weights were introduced in the fitting algorithm,
	wrong estimates of the parameters (the bulge parameters are
	the most affected ones) occurred; therefore, the
	equi-weighting scheme was chosen.

	The method described above was tested in several artificial
	galaxies. They were built with known and representative bulge
	and disk parameters. We chose typical profiles for this test,
	including: (a) those with well-defined bulge and disk, (b)
	with a dominant disk, (c) with a dominant bulge, (d) a nearly
	linear profile (fitted with a disk by our method) and (e) a
	curved profile (identified as a bulge by our method). The
	artificial profiles were convolved with a common seeing value
	of $1\farcs5$; the zone where this convolution was made was
	set randomly inside the typical interval of the true
	fits. Standard values of noise were added to the profile,
	based on real data. In Table \ref{test} some of the the input
	and output bulge and disk parameters are shown. The initial
	parameters seem to be well recovered by our technique; the
	largest differences correspond to profiles where the disk
	dominates although there is some contribution of a bulge
	component (test number 2, corresponding to a late-type
	spiral); these profiles were identified as an isolate disk by
	our method. Discrepancies were also present when one of the
	components is dominant (examples number 4 or 7, corresponding
	to a late-type spiral and an early-type galaxy, respectively);
	in this case, the parameters of the other component do not
	contribute much to the total profile and our method of
	decomposition does not recover the initial values (the errors
	of the B/D ratio are specially affected and are not shown in
	the result table -they are substituted by three dots-),
	although this fact is irrelevant. We took special care with
	these types of profile during morphological classification
	based on bulge-disk decomposition.

\begin{table*}
\caption{Bulge-disk decomposition test data}
\label{test}
\begin{tabular}{clllll}
\hline
\vspace{-0.3cm} & & \\
Profile type &  $\quad \mu_e$ & $\quad r_e$ & $\quad \mu_0$ & $\quad d_L$ & $\quad $B/D \\
\quad (1) & \quad (2) &\quad (3) & \quad (4) & \quad (5) & \quad (6)\\
\hline 
\hline 
a & 19.50          &  0.90          & 21.60          & 5.50           & 0.67
\\
 & 19.47$\pm$0.77 &  0.88$\pm$0.58 & 21.53$\pm$0.81 & 5.30$\pm$0.93  &
0.66$\pm$$0.36$\\
\hline
b & 26.00          & 13.80           & 21.50           & 4.60           &
0.51 \\
 & 25.34$\pm$0.23 &  5.95$\pm$4.17 & 21.32$\pm$0.41 & 4.54$\pm$1.42 &
0.15$\pm$0.33   \\
\hline
c & 23.30          &  5.30           & 21.80           & 6.90           &
0.54\\
 & 23.30$\pm$0.34 &  5.31$\pm$1.82 & 21.84$\pm$0.62 & 7.10$\pm$3.30 &
0.52$\pm$0.27   \\
\hline
d & 31.20           &  4.60           & 18.70           & 1.30           &
0.00\\
 & 26.76$\pm$0.04 &  1.60$\pm$0.99 & 18.71$\pm$0.01 & 1.30$\pm$0.02 &
0.00$\pm$0.01 \\
\hline
d & 24.10          &  4.40           & 19.20           & 2.40           &
0.13\\
 & 23.30$\pm$0.48 &  2.06$\pm$0.83 & 19.16$\pm$0.04 & 2.42$\pm$0.04 &
0.05$\pm$0.02   \\
\hline
e & 20.80          &  3.40          & 21.60          & 6.80          & 1.88\\
 & 20.80$\pm$0.13 &  3.40$\pm$0.39 & 21.62$\pm$0.74 & 6.94$\pm$2.81 &
1.84$\pm$0.51 \\
\hline
e & 21.50          &  3.50           & 22.90           & 4.90           &
6.68\\
 & 21.55$\pm$0.16 &  3.60$\pm$0.48 & 23.13$\pm$4.90 & 5.22$\pm$1.17 &
7.35$\pm$$\ldots$   \\
\hline
\hline
\end{tabular}
\setcounter{table}{0}
\caption{Results for the test of the bulge-disk decomposition procedure on
seven artificial galaxies. Input parameters are in the first row and
output results and their corresponding errors in the second
one. Columns: (1) Profile type as explained in the text. (2) Effective
surface brightness of the bulge in mag$\cdot$arcsec$^{-2}$ . (3)
Effective radius of the bulge in arcsec. (4) Typical surface
brightness of the disk in mag$\cdot$arcsec$^{-2}$. (5) Exponential
scale of the disk in arcsec. (6) Bulge-to-disk ratio}
\end{table*}

	One of the main problems during profile fitting is the fact
	that the hypersurface in the four parameters space
	($\mu_e$,$r_e$,$\mu_0$,$d_L$) has many local minima. The
	minimization method must be able to determine the real
	absolute minimum, whose parameters must have physical
	meaning. To achieve this, all the initial parameters were
	varied randomly before attempting the fit; we also used
	several fractional convergence tolerances in each individual
	fit and boundaries on each parameter were taken in order to
	avoid solutions with no physical meaning.

	With the four parameters of the disk-bulge decomposition, the
	bulge-to-disk luminosity ratio was calculated as follows:

	\begin{equation}
		\frac{B}{D}=\frac{L_B}{L_D}=3.607 \left ( \frac{r_e}{d_L}\right )^{2} \cdot 10^{(-0.4(\mu_e-\mu_0))}
	\end{equation}
	
	All the data referring to bulge-disk decomposition, along with
	mean ellipticities and position angles calculated as explained
	in Sect. \ref{data.1}, are shown in Table \ref{fe}. Some
	galaxies were unsuitable to perform bulge-disk decomposition
	due to very perturbed profiles or bad-quality of the
	images. These galaxies have no data of the bulge and disk
	parameters. Other galaxies were fitted with only one
	component; bulge-to-disk ratios for these objects have very
	large errors and no physical meaning so they are not shown in
	Table \ref{fe}. Position angles were only measured in those
	galaxies with ellipticities greater than 0.05 (for rounder
	isophotes, estimation of the PA is meaningless); two galaxies
	with $\epsilon$=0.06, but very large error values, have not PA
	measurement, either.

\begin{table*}
\tiny
{\normalsize \caption{ Bulge and disk parameters, bulge-to-disk ratio, ellipticity and position angle of the UCM Survey galaxies}}
\begin{tabular}{lcrccccr}
\hline
 {UCM name}& $\mu_e$ & r$_e$\,\,\,\,\,\,\,\,\,\,\,& $\mu_0$ & $d_L$ &  B/D &  $\epsilon$ & PA\,\,\,\,\,\\
 (1) & (2)  & (3)\,\,\,\,\,\,\,\,\,\, & (4) & (5) & (6) & (7) & (8)\,\,\,\,\,\\
\hline
\hline
0000+2140 & 18.96$\pm$0.47 &  0.69$\pm$0.70 & 19.74$\pm$0.38 &  4.15$\pm$0.68 &  0.20$\pm$0.24  & 0.28$\pm$0.01  &  $-$52$\pm$ 4 \\
0003+2200 & 21.04$\pm$0.38 &  0.07$\pm$0.56 & 21.35$\pm$0.04 &  4.73$\pm$0.40 &  0.00$\pm$0.00  & 0.59$\pm$0.04  &  $-$80$\pm$ 2 \\
0003+2215 & 27.71$\pm$2.73 & 21.19$\pm$4.72 & 21.71$\pm$0.36 &  6.04$\pm$2.60 &  0.18$\pm$0.13  & 0.59$\pm$0.04  & $-$2$\pm$ 3 \\
0003+1955 &       $-$        &$-$\,\,\,\,\,\,\,\,\,\,\,\,&       $-$        &       $-$        &       $-$         & 0.06$\pm$0.04  &$-$\,\,\,\,\,\\
0005+1802 & 21.31$\pm$0.35 &  0.80$\pm$0.39 & 20.57$\pm$0.09 &  4.27$\pm$0.47 &  0.06$\pm$0.03  & 0.64$\pm$0.01  &  76$\pm$ 1 \\
0006+2332 & 23.04$\pm$0.32 &  2.38$\pm$1.12 & 20.21$\pm$0.13 &  6.52$\pm$0.85 &  0.04$\pm$0.02  & 0.58$\pm$0.03  &  $-$22$\pm$ 1 \\
0013+1942 & 32.58$\pm$3.49 &  1.41$\pm$2.12 & 19.85$\pm$0.07 &  1.48$\pm$0.13 &  0.00$\pm$0.04  & 0.24$\pm$0.02  &      52$\pm$ 6 \\
0014+1829 & 24.86$\pm$1.81 & 14.96$\pm$3.84 & 18.85$\pm$0.26 &  1.02$\pm$0.08 &  3.06$\pm$1.42  & 0.30$\pm$0.06  &      54$\pm$ 6 \\
0014+1748 & 20.18$\pm$0.43 &  1.03$\pm$0.36 & 22.26$\pm$0.11 & 22.06$\pm$2.34\,\,\,&  0.05$\pm$0.01  & 0.72$\pm$0.01  &      46$\pm$ 1 \\
0015+2212 & 21.44$\pm$0.20 &  1.71$\pm$0.38 & 21.56$\pm$1.13 &  2.30$\pm$0.63 &  2.23$\pm$\ldots\,\,\,   & 0.09$\pm$0.05  &      12$\pm$14 \\
0017+1942 & 23.00$\pm$0.89 &  0.62$\pm$1.60 & 20.18$\pm$0.04 &  4.53$\pm$0.56 &  0.00$\pm$0.01  & 0.61$\pm$0.01  &   1$\pm$ 2 \\
0017+2148 & 27.83$\pm$0.50 &  0.81$\pm$0.46 & 19.05$\pm$0.05 &  1.23$\pm$0.06 &  0.00$\pm$0.00  & 0.22$\pm$0.04  &      40$\pm$ 4 \\
0018+2216 & 22.18$\pm$0.37 &  1.35$\pm$0.93 & 19.86$\pm$0.22 &  1.66$\pm$0.09 &  0.28$\pm$0.43  & 0.21$\pm$0.04  &  $-$72$\pm$ 3 \\
0018+2218 & 26.59$\pm$0.16 & 14.85$\pm$6.70 & 21.73$\pm$0.23 &  8.84$\pm$1.62 &  0.12$\pm$0.14  & 0.48$\pm$0.06  &  $-$66$\pm$ 1 \\
0019+2201 & 22.76$\pm$0.35 &  1.74$\pm$1.55 & 20.51$\pm$0.34 &  2.38$\pm$0.35 &  0.24$\pm$0.62  & 0.36$\pm$0.01  &   69$\pm$ 3 \\
0022+2049 & 23.44$\pm$0.93 &  1.26$\pm$1.69 & 19.75$\pm$0.14 &  3.56$\pm$0.37 &  0.02$\pm$0.04  & 0.51$\pm$0.02  &  $-$63$\pm$ 1 \\
0023+1908 &       $-$        &$-$\,\,\,\,\,\,\,\,\,\,\,\,&       $-$        &       $-$        &       $-$         & 0.19$\pm$0.06  &   40$\pm$ 3 \\
0034+2119 & 19.34$\pm$0.50 &  0.25$\pm$0.15 & 20.34$\pm$0.11 &  3.63$\pm$0.29 &  0.04$\pm$0.01  & 0.44$\pm$0.06  &  $-$85$\pm$ 2 \\
0037+2226 & 23.33$\pm$0.57 &  3.46$\pm$2.78 & 20.57$\pm$0.11 &  8.18$\pm$0.57 &  0.05$\pm$0.07  & 0.07$\pm$0.03  &  $-$22$\pm$11 \\
0038+2259 & 25.35$\pm$0.20 &  5.76$\pm$2.06 & 20.87$\pm$0.13 &  5.33$\pm$0.92 &  0.07$\pm$0.04  & 0.60$\pm$0.03  &   \,\,82$\pm$ 2 \\
0039+0054 &       $-$        &$-$\,\,\,\,\,\,\,\,\,\,\,\,&       $-$        &       $-$        &       $-$         & 0.31$\pm$0.07  &   23$\pm$ 5 \\
0040+0257 &       $-$        &$-$\,\,\,\,\,\,\,\,\,\,\,\,&       $-$        &       $-$        &       $-$         & 0.35$\pm$0.04  &  $-$26$\pm$ 4 \\
0040+2312 & 24.35$\pm$0.97 &  0.43$\pm$0.53 & 21.01$\pm$0.04 &  7.77$\pm$0.63 &  0.00$\pm$0.00  & 0.66$\pm$0.03  &   50$\pm$ 2 \\
0040+0220 & 23.59$\pm$3.20 &  1.29$\pm$1.28 & 19.29$\pm$0.13 &  1.20$\pm$0.07 &  0.08$\pm$0.07  & 0.11$\pm$0.05  &  $-$37$\pm$15 \\
0040$-$0023 & 24.27$\pm$0.19 &  7.35$\pm$1.75 & 20.13$\pm$0.05 &  8.32$\pm$0.64 &  0.06$\pm$0.03  & 0.24$\pm$0.00  &    6$\pm$ 0 \\
0041+0134 & 26.02$\pm$0.20 & 22.12$\pm$9.09 & 22.31$\pm$0.27 & 20.64$\pm$4.32\,\,\, &  0.14$\pm$0.18  & 0.23$\pm$0.00  &   70$\pm$ 0 \\
0043+0245 & 22.19$\pm$1.38 &  0.48$\pm$0.31 & 19.50$\pm$0.20 &  1.32$\pm$0.19 &  0.04$\pm$0.04  & 0.20$\pm$0.09  &   73$\pm$16 \\
0043$-$0159 & 19.17$\pm$0.23 &  1.23$\pm$0.21 & 21.31$\pm$0.11 & 20.71$\pm$1.26\,\,\, &  0.09$\pm$0.02  & 0.34$\pm$0.05  &  $-$30$\pm$ 6 \\
0044+2246 & 23.88$\pm$0.20 &  3.35$\pm$1.02 & 21.74$\pm$0.13 &  9.96$\pm$2.70 &  0.06$\pm$0.04  & 0.68$\pm$0.01  &   68$\pm$ 1 \\
0045+2206 &       $-$        &$-$\,\,\,\,\,\,\,\,\,\,\,\,&       $-$        &       $-$        &                    $-$         & 0.27$\pm$0.03  &      6$\pm$ 3 \\
0047+2051 & 20.60$\pm$0.45 &  0.28$\pm$0.29 & 20.05$\pm$0.10 &  1.95$\pm$0.19 &  0.04$\pm$0.01  & 0.17$\pm$0.02  &  $-$23$\pm$ 5 \\
0047$-$0213 & 21.25$\pm$0.26 &  2.63$\pm$0.53 & 22.32$\pm$1.06 &  8.82$\pm$2.33 &  0.86$\pm$\ldots\,\,\,   & 0.40$\pm$0.11  &   25$\pm$ 4 \\
0047+2413 & 22.26$\pm$0.32 &  2.15$\pm$1.16 & 21.34$\pm$0.30 &  6.20$\pm$0.43 &  0.19$\pm$0.22  & 0.51$\pm$0.04  &   40$\pm$ 4 \\
0047+2414 & 21.99$\pm$0.90 &  0.57$\pm$1.52 & 19.96$\pm$0.08 &  5.42$\pm$0.46 &  0.01$\pm$0.01  & 0.27$\pm$0.02  &  $-$74$\pm$ 4 \\
0049$-$0006 &       $-$        &$-$\,\,\,\,\,\,\,\,\,\,\,\,&       $-$        &       $-$        &       $-$         & 0.33$\pm$0.05  &  $-$76$\pm$ 5 \\
0049+0017 & 24.59$\pm$0.17 &  9.66$\pm$1.33 & 21.48$\pm$1.05 &  1.62$\pm$2.80 &  7.31$\pm$\ldots\,\,\,   & 0.40$\pm$0.05  &  $-$19$\pm$ 2 \\
0049$-$0045 &       $-$        &$-$\,\,\,\,\,\,\,\,\,\,\,\,&       $-$        &       $-$        &       $-$         &       $-$        & $-$\,\,\,\,\,     \\
0050+0005 & 23.12$\pm$0.27 &  4.66$\pm$1.72 & 20.61$\pm$0.66 &  2.73$\pm$0.97 &  1.04$\pm$2.21  & 0.41$\pm$0.03  &   73$\pm$ 3 \\
0050+2114 &       $-$        &$-$\,\,\,\,\,\,\,\,\,\,\,\,&       $-$        &       $-$        &       $-$         &        $-$       & $-$\,\,\,\,\,      \\
0051+2430 & 24.07$\pm$0.39 &  8.79$\pm$4.14 & 20.77$\pm$0.41 &  4.87$\pm$1.20 &  0.56$\pm$1.52  & 0.43$\pm$0.09  &   25$\pm$ 2 \\
0054$-$0133 & 20.89$\pm$0.72 &  0.31$\pm$1.21 & 20.16$\pm$0.20 &  3.47$\pm$0.75 &  0.01$\pm$0.06  & 0.36$\pm$0.03  &  $-$71$\pm$ 3 \\
0054+2337 & 24.02$\pm$0.40 &  5.39$\pm$3.36 & 20.99$\pm$0.33 & 10.85$\pm$2.12\,\,\, &  0.05$\pm$0.08  & 0.72$\pm$0.03  &   $-$7$\pm$ 2 \\
0056+0044 & 23.33$\pm$0.62 &  1.41$\pm$1.11 & 21.88$\pm$0.09 &  5.80$\pm$0.91 &  0.06$\pm$0.07  & 0.54$\pm$0.02  &  $-$30$\pm$ 2 \\
0056+0043 &       $-$        &$-$\,\,\,\,\,\,\,\,\,\,\,\,&       $-$        &       $-$        &       $-$         & 0.44$\pm$0.03  &   82$\pm$ 3 \\
0119+2156 & 22.63$\pm$0.37 &  0.82$\pm$1.15 & 20.86$\pm$0.15 &  3.63$\pm$0.59 &  0.04$\pm$0.05  & 0.68$\pm$0.03  &  42$\pm$ 2 \\
0121+2137 &       $-$        &$-$\,\,\,\,\,\,\,\,\,\,\,\,&       $-$        &       $-$        &       $-$         & 0.29$\pm$0.07  &   65$\pm$ 8 \\
0129+2109 & 22.51$\pm$0.66 &  1.23$\pm$1.15 & 20.95$\pm$0.17 &  6.48$\pm$1.30 &  0.03$\pm$0.03  & 0.18$\pm$0.04  &  $-$87$\pm$ 4 \\
0134+2257 & 20.58$\pm$0.34 &  0.53$\pm$0.34 & 21.09$\pm$0.16 &  4.20$\pm$1.27 &  0.09$\pm$0.06  & 0.11$\pm$0.04  &   23$\pm$ 0 \\
0135+2242 & 22.76$\pm$0.28 &  3.72$\pm$0.85 & 34.82$\pm$0.49 &  5.41$\pm$0.19 &   \ldots$\pm$\ldots   & 0.22$\pm$0.09  &   73$\pm$ 8 \\
0138+2216 &       $-$        &$-$\,\,\,\,\,\,\,\,\,\,\,\,&       $-$        &       $-$        &       $-$         & 0.55$\pm$0.02  &  $-$53$\pm$ 1 \\
0141+2220 & 23.28$\pm$0.34 &  2.61$\pm$1.55 & 19.90$\pm$0.14 &  3.21$\pm$0.19 &  0.11$\pm$0.11 &  0.64$\pm$0.03 &      39$\pm$ 2  \\
0142+2137 & 21.41$\pm$0.60 &  0.86$\pm$0.90 & 21.67$\pm$0.16 & 14.22$\pm$1.87\,\,\, &  0.02$\pm$0.02 &  0.69$\pm$0.05 &      36$\pm$ 4  \\
0144+2519 &       $-$        &$-$\,\,\,\,\,\,\,\,\,\,\,\,&       $-$        &       $-$        &       $-$        &  0.33$\pm$0.00 &      47$\pm$ 0  \\
0147+2309 & 25.21$\pm$0.04 & 16.29$\pm$0.00 & 21.09$\pm$0.01 &  2.61$\pm$0.02 &  3.16$\pm$0.12 &  0.54$\pm$0.03 &  $-$3$\pm$ 3  \\
0148+2124 & 23.51$\pm$0.16 &  3.38$\pm$0.45 & 21.10$\pm$0.48 &  1.82$\pm$0.70 &  1.35$\pm$1.05 &  0.20$\pm$0.11 &      3$\pm$17  \\
0150+2032 &       $-$        &$-$\,\,\,\,\,\,\,\,\,\,\,\,&       $-$        &       $-$        &       $-$        &  0.52$\pm$0.00 &      70$\pm$ 0  \\
0156+2410 &       $-$        &$-$\,\,\,\,\,\,\,\,\,\,\,\,&       $-$        &       $-$        &       $-$        &  0.50$\pm$0.03 &      73$\pm$ 2  \\
0157+2413 & 25.00$\pm$0.48 &  3.86$\pm$2.08 & 20.82$\pm$0.06 & 12.62$\pm$0.97\,\,\, &  0.01$\pm$0.00 &  0.77$\pm$0.01 &  $-$5$\pm$ 1  \\
0157+2102 &       $-$        &$-$\,\,\,\,\,\,\,\,\,\,\,\,&       $-$        &       $-$        &       $-$        &  0.58$\pm$0.04 &      83$\pm$ 3  \\
0159+2354 &       $-$        &$-$\,\,\,\,\,\,\,\,\,\,\,\,&       $-$        &       $-$        &       $-$        &  0.45$\pm$0.06 & $-$37$\pm$ 5  \\
0159+2326 & 22.18$\pm$0.47 &  0.49$\pm$0.58 & 20.04$\pm$0.08 &  3.26$\pm$0.23 &  0.01$\pm$0.01 &  0.29$\pm$0.02 &      32$\pm$ 3  \\
1246+2727 & 22.38$\pm$0.38 &  0.47$\pm$0.51 & 20.62$\pm$0.11 &  4.83$\pm$0.66 &  0.01$\pm$0.00 &  0.43$\pm$0.02 &  $-$5$\pm$ 2  \\
1247+2701 & 23.43$\pm$2.56 &  0.44$\pm$0.62 & 20.46$\pm$0.16 &  3.88$\pm$0.55 &  0.00$\pm$0.01 &  0.65$\pm$0.01 &      51$\pm$ 1  \\
1248+2912 & 20.89$\pm$0.49 &  0.61$\pm$0.85 & 20.39$\pm$0.12 &  4.75$\pm$0.78 &  0.04$\pm$0.04 &  0.32$\pm$0.02 & $-$71$\pm$ 3  \\
1253+2756 & 26.01$\pm$0.59 & 16.82$\pm$2.27 & 19.30$\pm$0.21 &  1.98$\pm$0.14 &  0.54$\pm$0.34 &  0.31$\pm$0.03 & $-$11$\pm$ 3  \\
1254+2741 &       $-$        &$-$\,\,\,\,\,\,\,\,\,\,\,\,&       $-$        &       $-$        &       $-$        &  0.58$\pm$0.02 & $-$65$\pm$ 2  \\
1254+2802 & 25.26$\pm$1.88 &  0.51$\pm$0.66 & 20.85$\pm$0.03 &  3.47$\pm$0.25 &  0.00$\pm$0.00 &  0.50$\pm$0.02 &      75$\pm$ 2  \\
1255+2819 &       $-$        &$-$\,\,\,\,\,\,\,\,\,\,\,\,&       $-$        &       $-$        &       $-$        &  0.13$\pm$0.04 &      50$\pm$10  \\
1255+3125 &       $-$        &$-$\,\,\,\,\,\,\,\,\,\,\,\,&       $-$        &       $-$        &       $-$        &  0.59$\pm$0.03 &      1$\pm$ 2  \\
1255+2734 &       $-$        &$-$\,\,\,\,\,\,\,\,\,\,\,\,&       $-$        &       $-$        &       $-$        &  0.49$\pm$0.06 & $-$35$\pm$ 2  \\
1256+2717 & 23.94$\pm$0.38 &  3.82$\pm$0.53 & 20.70$\pm$0.57 &  1.07$\pm$0.27 &  2.33$\pm$5.18 &  0.30$\pm$0.03 &      54$\pm$ 4  \\
1256+2732 & 21.43$\pm$0.50 &  1.73$\pm$1.15 & 21.17$\pm$0.72 &  4.12$\pm$1.46 &  0.50$\pm$0.45 &  0.34$\pm$0.06 &      82$\pm$ 5  \\
1256+2701 & 22.59$\pm$0.68 &  0.46$\pm$0.78 & 21.71$\pm$0.09 &  9.29$\pm$2.25 &  0.00$\pm$0.00 &  0.80$\pm$0.01 & $-$64$\pm$ 1  \\
1256+2910 & 22.01$\pm$0.38 &  1.71$\pm$1.96 & 21.47$\pm$0.75 &  4.41$\pm$0.74 &  0.33$\pm$2.58 &  0.15$\pm$0.02 &      21$\pm$ 7  \\
1256+2823 &       $-$        &$-$\,\,\,\,\,\,\,\,\,\,\,\,&       $-$        &       $-$        &       $-$        &  0.19$\pm$0.01 &      31$\pm$ 5  \\
1256+2754 & 22.50$\pm$0.37 &  6.89$\pm$2.20 & 25.28$\pm$4.58 &  6.36$\pm$2.20 & 54.79$\pm$\ldots\,\,\,\,\,\,  &  0.15$\pm$0.06 & $-$80$\pm$10  \\
1256+2722 & 23.23$\pm$0.12 &  0.76$\pm$0.16 & 20.46$\pm$0.09 &  3.35$\pm$0.39 &  0.01$\pm$0.00 &  0.64$\pm$0.01 & $-$48$\pm$ 1  \\
1257+2808 & 25.76$\pm$0.48 & 13.13$\pm$4.87 & 19.57$\pm$0.10 &  1.96$\pm$0.13 &  0.54$\pm$0.42 &  0.29$\pm$0.02 &  $-$6$\pm$ 2  \\
1258+2754 & 21.50$\pm$0.76 &  1.12$\pm$2.79 & 20.71$\pm$0.29 &  4.53$\pm$0.61 &  0.11$\pm$0.45 &  0.37$\pm$0.07 &      58$\pm$ 4  \\
1259+2934 &       $-$        &$-$\,\,\,\,\,\,\,\,\,\,\,\,&       $-$        &       $-$        &       $-$        &        $-$       & $-$\,\,\,\,\,  \\
1259+3011 & 20.29$\pm$0.30 &  1.14$\pm$0.32 & 21.22$\pm$0.47 &  3.72$\pm$0.80 &  0.80$\pm$1.02 &  0.38$\pm$0.02 &      30$\pm$ 2  \\
1259+2755 & 23.72$\pm$0.27 &  7.82$\pm$3.83 & 19.70$\pm$0.15 &  3.37$\pm$0.22 &  0.48$\pm$0.71 &  0.45$\pm$0.05 &      87$\pm$ 5  \\
1300+2907 & 22.96$\pm$0.31 &  3.18$\pm$0.85 & 20.92$\pm$0.37 &  2.31$\pm$0.46 &  1.16$\pm$0.66 &  0.51$\pm$0.03 &      54$\pm$ 2  \\
1301+2904 & 25.11$\pm$0.55 &  8.96$\pm$5.73 & 20.88$\pm$0.19 &  4.49$\pm$0.39 &  0.29$\pm$0.41 &  0.13$\pm$0.02 &      16$\pm$ 5  \\
1302+2853 & 21.98$\pm$0.95 &  0.12$\pm$0.16 & 19.66$\pm$0.07 &  2.40$\pm$0.11 &  0.00$\pm$0.01 &  0.40$\pm$0.02 &      85$\pm$ 5  \\
1302+3032 & 24.78$\pm$0.70 &  9.06$\pm$5.99 & 19.68$\pm$0.17 &  1.55$\pm$0.18 &  1.12$\pm$1.51 &  0.34$\pm$0.02 & $-$55$\pm$ 2  \\
1303+2908 & 25.14$\pm$2.67 &  0.31$\pm$1.41 & 21.35$\pm$0.06 &  4.79$\pm$0.54 &  0.00$\pm$0.00 &  0.57$\pm$0.04 &      8$\pm$ 3  \\
1304+2808 & 28.17$\pm$0.10 &  1.35$\pm$0.43 & 20.52$\pm$0.01 &  4.33$\pm$0.11 &  0.00$\pm$0.00 &  0.49$\pm$0.04 & $-$48$\pm$ 1  \\
1304+2830 & 24.87$\pm$2.86 &  0.28$\pm$0.95 & 19.79$\pm$0.15 &  0.82$\pm$0.04 &  0.00$\pm$0.17 &  0.18$\pm$0.02 &      86$\pm$ 4  \\
1304+2907 & 25.76$\pm$0.87 & 18.14$\pm$1.25 & 21.73$\pm$0.50 &  8.62$\pm$0.60 &  0.39$\pm$0.44 &  0.20$\pm$0.00 &      6$\pm$ 0  \\
1304+2818 & 23.74$\pm$0.53 &  1.31$\pm$1.27 & 20.65$\pm$0.08 &  3.98$\pm$0.39 &  0.02$\pm$0.02 &  0.14$\pm$0.02 &      54$\pm$ 2  \\
1306+2938 & 25.65$\pm$0.35 & 19.09$\pm$4.05 & 19.09$\pm$0.09 &  2.46$\pm$0.26 &  0.52$\pm$0.38 &  0.18$\pm$0.07 & $-$18$\pm$11  \\
1306+3111 & 22.70$\pm$2.41 &  0.16$\pm$0.38 & 19.92$\pm$0.04 &  2.26$\pm$0.12 &  0.00$\pm$0.00 &  0.19$\pm$0.03 & $-$63$\pm$ 5  \\
1307+2910 & 21.48$\pm$0.34 &  1.50$\pm$0.73 & 21.15$\pm$0.16 & 15.72$\pm$3.58\,\,\, &  0.02$\pm$0.01 &  0.57$\pm$0.02 &      86$\pm$ 1  \\
1308+2958 & 21.22$\pm$0.35 &  0.75$\pm$0.66 & 21.58$\pm$0.15 &  8.13$\pm$1.81 &  0.04$\pm$0.02 &  0.36$\pm$0.04 &      0$\pm$ 4  \\
1308+2950 &       $-$        &$-$\,\,\,\,\,\,\,\,\,\,\,\,&       $-$        &       $-$        &       $-$        &  0.65$\pm$0.02 & $-$11$\pm$ 2  \\
1310+3027 & 24.21$\pm$0.27 &  5.13$\pm$1.65 & 20.53$\pm$0.23 &  3.11$\pm$0.43 &  0.33$\pm$0.17 &  0.56$\pm$0.02 & $-$31$\pm$ 2  \\
1312+3040 & 21.58$\pm$0.69 &  1.50$\pm$1.40 & 20.32$\pm$0.45 &  3.73$\pm$0.66 &  0.18$\pm$0.26 &  0.35$\pm$0.09 & $-$84$\pm$ 3  \\
1312+2954 &       $-$        &$-$\,\,\,\,\,\,\,\,\,\,\,\,&       $-$        &       $-$        &                &  0.62$\pm$0.06 & $-$82$\pm$ 1  \\
1313+2938 & 22.20$\pm$0.15 &  2.06$\pm$1.62 & 18.59$\pm$0.21 &  0.95$\pm$0.05 &  0.61$\pm$1.07 &        $-$     & $-$\,\,\,\,\,   \\
\hline										     
\hline										     
\end{tabular}\\								     
\end{table*}									     
\setcounter{table}{1}							     
\begin{table*}									     
\tiny
{\normalsize \caption{continued}}
\begin{tabular}{lcrccccr}
\hline
{UCM name}& $\mu_e$ & r$_e$\,\,\,\,\,\,\,\,\,\,\, & $\mu_0$ & $d_L$ &  B/D &  $\epsilon$ & PA\,\,\,\,\,\\
 (1) & (2)  & (3)\,\,\,\,\,\,\,\,\,\, & (4) & (5) & (6) & (7) & (8)\,\,\,\,\,\\
\hline
\hline
1314+2827 &       $-$        &$-$\,\,\,\,\,\,\,\,\,\,\,\,&       $-$        &       $-$        &       $-$        & 0.01$\pm$0.04 &      $-$\,\,\,\,\,      \\
1320+2727 & 23.94$\pm$0.88 &  4.41$\pm$0.59 & 19.73$\pm$0.22 &  1.10$\pm$0.12 &  1.20$\pm$0.13 & 0.32$\pm$0.05 &      75$\pm$ 5  \\
1324+2926 & 24.91$\pm$0.45 &  5.85$\pm$0.47 & 19.62$\pm$0.21 &  0.70$\pm$0.04 &  1.93$\pm$1.23 & 0.11$\pm$0.04 &      54$\pm$21  \\
1324+2651 & 17.68$\pm$0.52 &  0.53$\pm$0.23 & 21.12$\pm$0.51 &  5.36$\pm$1.25 &  0.84$\pm$0.52 & 0.33$\pm$0.01 & $-$24$\pm$ 1  \\
1331+2900 &       $-$        &$-$\,\,\,\,\,\,\,\,\,\,\,\,&       $-$        &       $-$        &       $-$        & 0.25$\pm$0.14 & $-$23$\pm$18  \\
1428+2727 & 25.90$\pm$3.09 &  2.85$\pm$2.34 & 19.22$\pm$0.11 &  3.96$\pm$0.30 &  0.00$\pm$0.02 & 0.45$\pm$0.01 &      83$\pm$ 1  \\
1429+2645 & 21.33$\pm$0.10 &  0.32$\pm$0.12 & 19.91$\pm$0.20 &  1.14$\pm$0.12 &  0.08$\pm$0.08 & 0.11$\pm$0.04 & $-$41$\pm$18  \\
1430+2947 & 22.84$\pm$0.26 &  4.45$\pm$0.80 & 25.35$\pm$2.92 & 11.47$\pm$4.24\,\,\, &  5.48$\pm$\ldots\,\,\,  & 0.23$\pm$0.04 & $-$87$\pm$ 4  \\
1431+2854 & 24.16$\pm$0.36 &  8.65$\pm$3.92 & 20.38$\pm$0.21 &  3.36$\pm$0.32 &  0.74$\pm$0.73 & 0.21$\pm$0.04 & $-$68$\pm$ 5  \\
1431+2702 & 25.12$\pm$0.47 & 10.05$\pm$2.03 & 18.85$\pm$0.13 &  0.77$\pm$0.05 &  1.91$\pm$0.67 & 0.10$\pm$0.06 &      67$\pm$10  \\
1431+2947 &       $-$        &$-$\,\,\,\,\,\,\,\,\,\,\,\,&       $-$        &       $-$        &       $-$        & 0.39$\pm$0.03 & $-$50$\pm$ 5  \\
1431+2814 & 23.38$\pm$0.19 &  2.30$\pm$1.57 & 20.41$\pm$0.17 &  2.93$\pm$0.23 &  0.14$\pm$0.31 & 0.65$\pm$0.01 & $-$25$\pm$ 1  \\
1432+2645 & 20.84$\pm$0.88 &  1.23$\pm$1.09 & 22.00$\pm$0.83 & 10.04$\pm$3.64\,\,\, &  0.16$\pm$0.20 & 0.42$\pm$0.01 & $-$78$\pm$ 6  \\
1440+2521S&       $-$        &$-$\,\,\,\,\,\,\,\,\,\,\,\,&       $-$        &       $-$        &       $-$        & 0.41$\pm$0.03 & 24$\pm$ 2  \\
1440+2511 & 21.70$\pm$0.93 &  1.04$\pm$0.83 & 22.55$\pm$0.38 &  8.15$\pm$3.20 &  0.13$\pm$0.07 & 0.31$\pm$0.09 & $-$24$\pm$ 3  \\
1440+2521N& 23.52$\pm$0.29 &  3.00$\pm$1.38 & 21.42$\pm$0.22 &  4.53$\pm$0.52 &  0.23$\pm$0.15 & 0.41$\pm$0.05 &      62$\pm$ 9  \\
1442+2845 & 25.64$\pm$0.29 & 21.69$\pm$7.73 & 20.34$\pm$0.13 &  3.25$\pm$0.16 &  1.22$\pm$1.05 & 0.06$\pm$0.01 & $-$33$\pm$ 6  \\
1443+2714 & 22.14$\pm$0.13 &  3.29$\pm$0.52 & 21.24$\pm$0.35 &  4.24$\pm$0.78 &  0.95$\pm$0.37 & 0.12$\pm$0.04 &      72$\pm$24  \\
1443+2844 & 23.72$\pm$2.49 &  0.13$\pm$1.10 & 20.36$\pm$0.02 &  3.98$\pm$0.38 &  0.00$\pm$0.00 & 0.49$\pm$0.01 & $-$58$\pm$ 1  \\
1443+2548 & 22.94$\pm$0.24 &  1.19$\pm$0.64 & 20.29$\pm$0.09 &  3.88$\pm$0.37 &  0.03$\pm$0.02 & 0.21$\pm$0.03 & $-$82$\pm$ 8  \\
1444+2923 & 21.90$\pm$0.58 &  1.84$\pm$1.15 & 23.11$\pm$0.83 &  7.06$\pm$2.57 &  0.75$\pm$1.08 & 0.12$\pm$0.06 & $-$13$\pm$11  \\
1452+2754 & 22.55$\pm$0.27 &  2.49$\pm$0.74 & 21.08$\pm$0.23 &  3.50$\pm$0.48 &  0.47$\pm$0.24 & 0.51$\pm$0.01 & $-$44$\pm$ 1  \\
1506+1922 & 20.29$\pm$0.27 &  0.71$\pm$0.25 & 21.21$\pm$0.12 &  5.22$\pm$0.33 &  0.16$\pm$0.03 & 0.43$\pm$0.01 &      68$\pm$ 1  \\
1513+2012 & 27.47$\pm$0.44 &  1.22$\pm$0.99 & 19.09$\pm$0.03 &  2.64$\pm$0.09 &  0.00$\pm$0.00 & 0.50$\pm$0.02 &      55$\pm$ 2  \\
1537+2506N& 19.39$\pm$0.37 &  0.74$\pm$0.19 & 20.95$\pm$0.18 &  5.91$\pm$0.57 &  0.24$\pm$0.08 & 0.35$\pm$0.05 & $-$69$\pm$ 5  \\
1537+2506S& 27.68$\pm$0.41 & 74.55$\pm$9.86 & 19.49$\pm$0.11 &  2.06$\pm$0.30 &  2.50$\pm$1.96 & 0.44$\pm$0.04 &      64$\pm$ 9  \\
1557+1423 & 24.84$\pm$3.84 &  1.60$\pm$1.01 & 20.06$\pm$0.12 &  2.13$\pm$0.13 &  0.02$\pm$0.05 & 0.21$\pm$0.02 &      35$\pm$ 3  \\
1612+1308 &       $-$        &$-$\,\,\,\,\,\,\,\,\,\,\,\,&       $-$        &       $-$        &       $-$        & 0.03$\pm$0.05 &      $-$\,\,\,\,\,      \\
1646+2725 &       $-$        &$-$\,\,\,\,\,\,\,\,\,\,\,\,&       $-$        &       $-$        &       $-$        & 0.68$\pm$0.02 & $-$70$\pm$ 1  \\
1647+2950 & 20.98$\pm$0.44 &  0.89$\pm$0.74 & 20.41$\pm$0.23 &  3.85$\pm$0.52 &  0.11$\pm$0.11 & 0.20$\pm$0.01 & $-$33$\pm$ 2  \\
1647+2729 & 22.32$\pm$0.18 &  0.52$\pm$0.55 & 20.10$\pm$0.10 &  3.49$\pm$0.29 &  0.01$\pm$0.01 & 0.43$\pm$0.02 &      89$\pm$ 2  \\
1647+2727 &       $-$        &$-$\,\,\,\,\,\,\,\,\,\,\,\,&       $-$        &       $-$        &       $-$        & 0.44$\pm$0.01 & $-$90$\pm$ 1  \\
1648+2855 & 25.35$\pm$0.46 & 15.02$\pm$3.16 & 19.52$\pm$0.34 &  2.08$\pm$0.18 &  0.88$\pm$1.07 & 0.16$\pm$0.04 &      3$\pm$ 4  \\
1653+2644 & 21.60$\pm$0.71 &  2.65$\pm$2.11 & 19.19$\pm$0.24 &  2.94$\pm$0.20 &  0.32$\pm$0.57 & 0.20$\pm$0.01 &   1$\pm$ 2  \\
1654+2812 & 26.69$\pm$3.87 &  0.88$\pm$1.47 & 21.79$\pm$0.06 &  2.96$\pm$0.36 &  0.00$\pm$0.01 & 0.60$\pm$0.02 & $-$49$\pm$ 2  \\
1655+2755 & 24.43$\pm$0.16 &  6.48$\pm$2.06 & 21.95$\pm$0.13 & 10.83$\pm$1.67\,\,\, &  0.13$\pm$0.07 & 0.49$\pm$0.10 &      38$\pm$10  \\
1656+2744 &       $-$        &$-$\,\,\,\,\,\,\,\,\,\,\,\,&       $-$        &       $-$        &       $-$        & 0.48$\pm$0.09 &      27$\pm$ 7  \\
1657+2901 &       $-$        &$-$\,\,\,\,\,\,\,\,\,\,\,\,&       $-$        &       $-$        &       $-$        & 0.51$\pm$0.01 &      82$\pm$ 1  \\
1659+2928 & 19.76$\pm$0.32 &  0.93$\pm$0.30 & 21.64$\pm$0.39 &  5.47$\pm$1.45 &  0.59$\pm$0.15 & 0.31$\pm$0.04 & $-$71$\pm$ 2  \\
1701+3131 & 19.24$\pm$0.16 &  0.89$\pm$0.24 & 21.41$\pm$0.35 &  6.85$\pm$2.53 &  0.45$\pm$0.24 & 0.32$\pm$0.10 &      59$\pm$ 8  \\
2238+2308 & 23.89$\pm$0.14 & 13.24$\pm$2.45 & 21.30$\pm$0.69 &  6.16$\pm$2.60 &  1.53$\pm$1.78 & 0.20$\pm$0.03 &  $-$6$\pm$ 4  \\
2239+1959 & 21.06$\pm$0.21 &  4.00$\pm$0.47 & 21.47$\pm$0.47 &  4.95$\pm$1.26 &  3.44$\pm$0.71 & 0.35$\pm$0.02 &      38$\pm$ 3  \\
2249+2149 & 22.44$\pm$0.27 &  3.41$\pm$0.63 & 22.04$\pm$0.35 &  7.42$\pm$2.37 &  0.53$\pm$0.15 & 0.54$\pm$0.03 & 58$\pm$ 2  \\
2250+2427 & 17.65$\pm$0.14 &  0.47$\pm$0.04 & 21.40$\pm$0.09 & 10.50$\pm$0.15\,\,\, &  0.23$\pm$0.03 & 0.51$\pm$0.06 & $-$21$\pm$ 2  \\
2251+2352 & 21.31$\pm$3.85 &  0.11$\pm$1.37 & 18.71$\pm$0.05 &  1.27$\pm$0.05 &  0.00$\pm$0.01 & 0.02$\pm$0.03 &      $-$\,\,\,\,\,      \\
2253+2219 & 22.48$\pm$1.09 &  1.09$\pm$1.56 & 19.11$\pm$0.12 &  2.31$\pm$0.10 &  0.04$\pm$0.08 & 0.56$\pm$0.02 & 32$\pm$ 1  \\
2255+1930S& 21.11$\pm$1.48 &  0.43$\pm$0.53 & 18.51$\pm$0.26 &  1.37$\pm$0.13 &  0.03$\pm$0.14 & 0.30$\pm$0.05 &      41$\pm$ 4  \\
2255+1930N& 24.35$\pm$0.20 &  6.98$\pm$2.63 & 19.87$\pm$0.15 &  3.29$\pm$0.20 &  0.26$\pm$0.28 & 0.55$\pm$0.01 & $-$87$\pm$ 1  \\
2255+1926 & 26.24$\pm$0.20 & 15.81$\pm$2.28 & 21.36$\pm$0.19 &  3.89$\pm$0.66 &  0.67$\pm$0.18 & 0.57$\pm$0.03 & $-$17$\pm$ 3  \\
2255+1654 & 26.26$\pm$0.23 &  9.89$\pm$5.32 & 21.45$\pm$0.17 &  8.22$\pm$1.60 &  0.06$\pm$0.08 & 0.76$\pm$0.01 &      73$\pm$ 1  \\
2256+2001 & 23.66$\pm$0.64 &  1.49$\pm$1.42 & 22.23$\pm$0.14 & 10.81$\pm$2.03\,\,\, &  0.02$\pm$0.01 & 0.20$\pm$0.02  & $-$5$\pm$ 2  \\
2257+2438 & 19.31$\pm$0.09 &  1.10$\pm$0.06 & 27.33$\pm$3.30 &  2.80$\pm$0.71 &   \ldots$\pm$\ldots  & 0.20$\pm$0.03  & $-$55$\pm$ 5  \\
2257+1606 & 20.44$\pm$0.36 &  1.07$\pm$0.43 & 21.09$\pm$1.16 &  2.35$\pm$1.25 &  1.36$\pm$\ldots\,\,\,  & 0.06$\pm$0.07  &      $-$\,\,\,\,\,      \\
2258+1920 &       $-$        &$-$\,\,\,\,\,\,\,\,\,\,\,\,&       $-$        &       $-$        &       $-$        & 0.23$\pm$0.04  &      72$\pm$ 5  \\
2300+2015 &       $-$        &$-$\,\,\,\,\,\,\,\,\,\,\,\,&       $-$        &       $-$        &       $-$        & 0.08$\pm$0.03  & $-$64$\pm$14  \\
2302+2053W& 23.47$\pm$0.38 &  3.23$\pm$1.74 & 22.40$\pm$1.12 &  5.13$\pm$1.39 &  0.53$\pm$\ldots\,\,\,  & 0.62$\pm$0.01  &      66$\pm$ 1  \\
2302+2053E& 23.33$\pm$0.17 &  5.17$\pm$0.85 & 21.76$\pm$0.29 &  6.92$\pm$1.19 &  0.47$\pm$0.23 & 0.30$\pm$0.02  &      32$\pm$ 9  \\
2303+1856 &       $-$        &$-$\,\,\,\,\,\,\,\,\,\,\,\,&       $-$        &       $-$        &       $-$        & 0.41$\pm$0.07  & $-$24$\pm$ 7  \\
2303+1702 & 24.79$\pm$0.57 &  2.10$\pm$0.62 & 21.24$\pm$0.19 &  2.70$\pm$0.49 &  0.08$\pm$0.02 & 0.21$\pm$0.08  &      48$\pm$ 3  \\
2304+1640 &       $-$        &$-$\,\,\,\,\,\,\,\,\,\,\,\,&       $-$        &       $-$        &       $-$        & 0.23$\pm$0.07  & $-$50$\pm$14  \\
2304+1621 & 20.30$\pm$0.03 &  0.56$\pm$0.02 & 21.21$\pm$0.04 &  2.86$\pm$0.01 &  0.32$\pm$0.02 & 0.31$\pm$0.07  &  $-$6$\pm$ 5  \\
2307+1947 & 20.50$\pm$0.65 &  0.65$\pm$0.97 & 21.07$\pm$0.82 &  3.33$\pm$0.46 &  0.23$\pm$2.01 & 0.43$\pm$0.01  & $-$85$\pm$ 1  \\
2310+1800 & 21.45$\pm$0.26 &  0.59$\pm$1.02 & 20.39$\pm$0.23 &  2.49$\pm$0.23 &  0.08$\pm$0.24 & 0.32$\pm$0.15  &      14$\pm$11  \\
2312+2204 & 21.11$\pm$0.23 &  0.61$\pm$0.78 & 20.70$\pm$0.56 &  2.15$\pm$0.79 &  0.20$\pm$2.80 & 0.37$\pm$0.04  & $-$24$\pm$ 4  \\
2313+1841 &       $-$        &$-$\,\,\,\,\,\,\,\,\,\,\,\,&       $-$        &       $-$        &       $-$        & 0.57$\pm$0.04  & $-$33$\pm$ 4  \\
2313+2517 & 20.93$\pm$0.19 &  1.37$\pm$0.45 & 20.07$\pm$0.15 &  4.49$\pm$0.38 &  0.15$\pm$0.08 & 0.36$\pm$0.02  & 7$\pm$ 1  \\
2315+1923 & 19.48$\pm$0.12 &  0.23$\pm$0.12 & 20.62$\pm$0.12 &  2.15$\pm$0.02 &  0.12$\pm$0.05 & 0.50$\pm$0.06  &  $-$4$\pm$ 5  \\
2316+2457 & 21.08$\pm$0.17 &  3.53$\pm$0.31 & 20.95$\pm$0.31 &  5.79$\pm$0.22 &  1.19$\pm$0.31 & 0.02$\pm$0.02  &      $-$\,\,\,\,\,      \\
2316+2459 &       $-$        &$-$\,\,\,\,\,\,\,\,\,\,\,\,&       $-$        &       $-$        &       $-$        & 0.26$\pm$0.03  &      28$\pm$ 4  \\
2316+2028 &       $-$        &$-$\,\,\,\,\,\,\,\,\,\,\,\,&       $-$        &       $-$        &       $-$        & 0.28$\pm$0.06  &      32$\pm$12  \\
2317+2356 & 22.11$\pm$0.53 &  5.17$\pm$3.13 & 20.51$\pm$0.48 &  9.03$\pm$2.56 &  0.27$\pm$0.34 & 0.22$\pm$0.03  & $-$73$\pm$ 5  \\
2319+2234 & 21.97$\pm$1.17 &  0.47$\pm$0.61 & 19.54$\pm$0.15 &  1.86$\pm$0.25 &  0.02$\pm$0.02 & 0.41$\pm$0.04  &      88$\pm$ 3  \\
2319+2243 & 22.38$\pm$0.41 &  3.35$\pm$1.43 & 20.89$\pm$0.52 &  4.04$\pm$1.16 &  0.63$\pm$0.47 & 0.41$\pm$0.08  & $-$29$\pm$ 4  \\
2320+2428 & 21.60$\pm$0.49 &  0.41$\pm$0.55 & 20.11$\pm$0.08 &  4.59$\pm$0.55 &  0.01$\pm$0.01 & 0.73$\pm$0.03  &     8$\pm$ 1  \\
2321+2149 & 23.77$\pm$0.50 &  1.40$\pm$1.32 & 20.19$\pm$0.26 &  2.30$\pm$0.38 &  0.05$\pm$0.10 & 0.12$\pm$0.02  & $-$51$\pm$ 7  \\
2321+2506 & 21.51$\pm$0.78 &  0.25$\pm$0.89 & 20.73$\pm$0.07 &  5.82$\pm$0.42 &  0.00$\pm$0.01 & 0.50$\pm$0.02  &      60$\pm$ 5  \\
2322+2218 & 22.59$\pm$0.87 &  0.90$\pm$0.72 & 21.01$\pm$0.19 &  3.08$\pm$0.48 &  0.07$\pm$0.04 & 0.56$\pm$0.01  & $-$31$\pm$ 1  \\
2324+2448 & 22.57$\pm$0.49 &  2.40$\pm$2.54 & 20.07$\pm$0.10 & 12.57$\pm$1.15\,\,\, &  0.01$\pm$0.02 & 0.58$\pm$0.01  & $-$83$\pm$ 1  \\
2325+2318 &       $-$        &$-$\,\,\,\,\,\,\,\,\,\,\,\,&       $-$        &       $-$        &       $-$        &       $-$        &      $-$\,\,\,\,\,      \\
2325+2208 &       $-$        &$-$\,\,\,\,\,\,\,\,\,\,\,\,&       $-$        &       $-$        &       $-$        & 0.18$\pm$0.03  &      36$\pm$ 1  \\
2326+2435 &       $-$        &$-$\,\,\,\,\,\,\,\,\,\,\,\,&       $-$        &       $-$        &       $-$        & 0.74$\pm$0.02  & $-$73$\pm$ 1  \\
2327+2515N& 20.77$\pm$0.46 &  0.97$\pm$0.92 & 19.07$\pm$0.20 &  2.45$\pm$0.24 &  0.12$\pm$0.15 & 0.43$\pm$0.08  &  $-$1$\pm$ 5  \\
2327+2515S& 21.00$\pm$0.44 &  1.83$\pm$1.03 & 21.53$\pm$0.64 &  5.94$\pm$2.68 &  0.56$\pm$9.90 & 0.20$\pm$0.05  &      33$\pm$ 5  \\
2329+2427 & 21.31$\pm$0.22 &  0.86$\pm$0.26 & 21.29$\pm$0.14 &  6.20$\pm$1.30 &  0.07$\pm$0.03 & 0.62$\pm$0.03  & $-$58$\pm$ 1  \\
2329+2500 & 19.83$\pm$0.23 &  0.88$\pm$0.18 & 23.17$\pm$0.52 & 13.92$\pm$0.75\,\,\, &  0.31$\pm$0.56 & 0.36$\pm$0.04  & $-$65$\pm$ 2  \\
2329+2512 & 24.86$\pm$2.67 &  6.47$\pm$0.43 & 19.90$\pm$0.22 &  1.17$\pm$0.07 &  1.14$\pm$0.64 & 0.26$\pm$0.07  &      70$\pm$ 6  \\
2331+2214 & 21.74$\pm$0.42 &  0.76$\pm$0.41 & 21.95$\pm$0.29 &  3.36$\pm$0.73 &  0.22$\pm$0.15 & 0.43$\pm$0.04  &      14$\pm$ 1  \\
2333+2248 & 21.52$\pm$0.79 &  0.15$\pm$0.58 & 21.54$\pm$0.08 &  5.02$\pm$0.85 &  0.00$\pm$0.00 & 0.72$\pm$0.04  & $-$64$\pm$ 3  \\
2333+2359 & 24.09$\pm$0.24 &  4.87$\pm$1.12 & 23.27$\pm$1.72 &  1.65$\pm$1.05 & 14.77$\pm$\ldots\,\,\,\,\,  & 0.12$\pm$0.08  & $-$20$\pm$21  \\
2348+2407 &       $-$        &$-$\,\,\,\,\,\,\,\,\,\,\,\,&       $-$        &       $-$        &       $-$        & 0.28$\pm$0.05  &      25$\pm$ 7  \\
2351+2321 & 21.51$\pm$0.32 &  0.57$\pm$0.34 & 21.60$\pm$0.26 &  1.40$\pm$0.39 &  0.65$\pm$4.44 & 0.13$\pm$0.05  &      61$\pm$11  \\
\hline \hline
\label{fe}
\end{tabular}\\
\vspace{0.5cm}
\setcounter{table}{1}
{\normalsize \caption{(1) UCM name. (2) Effective surface brightness of the bulge in mag$\cdot$arcsec$^{-2}$. (3) Scale of the bulge in arcsec. (4)
	Characteristic surface brightness of the disk in
	mag$\cdot$arcsec$^{-2}$. (5) Exponential disk scale in
	arcsec. (6) Bulge-to-disk ratio. (7) Ellipticity of the galaxy
	calculated as an average of the ellipticities of the isophotes
	between 23 and 24 mag$\cdot$arcsec$^{-2}$. (8) Position angle
	in degrees measured counterclockwise from the North axis and
	calculated as an average of the previously detailed isophotes}}
\end{table*}

\subsection{Light concentration indices}
	
	In order to characterize galaxies morphologically, we need a
	measure of the degree of concentration of light towards the
	central or the outermost regions of the object. Traditionally,
	this has been performed through the calculation of the
	bulge-to-disk ratio B/D mentioned in the previous Section. The
	methodology and degree of fidelity of the bulge-disk
	decomposition are trustful just for galaxy profiles where only
	these two components are present and no special features are
	found (Schombert \& Bothun \cite{Sch87}). As was already
	mentioned, some of our galaxies are dominated by bright
	features (mainly bars, spiral arms and bright starbursts) and
	it seems better to improve their morphological classification
	using concentration indices not based in previous component
	fits.

	We have calculated for the whole sample three concentration
	indices in the same way that Vitores et al. (\cite{Alv96a})
	did for the Gunn $r$ bandpass:

	\begin{itemize} \item c$_{31}$, defined by de Vaucouleurs (\cite{Vau77}) as the
	ratio between the radius containing 75\% of the light of the
	galaxy ($r_{75}$) and the radius with 25\% of the galaxy
	luminosity ($r_{25}$):

	\begin{equation} \label{31} c_{31}={r_{75} \over r_{25}}
	\end{equation}
	
	The calculation was performed integrating the flux of the
	galaxy in contiguous isophotes until 25 and 75\% of the total
	flux was reached. The radii used in Eq. (\ref{31}) are
	equivalent radii, calculated as the square root of the product
	of the semi-major and semi-minor axes. This index was found to
	correlate well with morphological type, decreasing from early
	to late-type galaxies (Gavazzi et al. \cite{Gav90}).

	\item c$_{42}$, defined by Kent (\cite{Ke85}) as:

		\begin{equation} \label{42} c_{42}=5\,{log\left ( r_{80} \over r_{20} \right )}
	\end{equation}
	
	where $r_{80}$ and $r_{20}$ are the equivalent radii
	containing 80 and 20\% of the total flux of the galaxy,
	respectively. These radii were calculated as the ones for
	c$_{31}$.  
	
	\item $c_{in}(\alpha)$, as defined in Doi et al. (\cite{Doi93}):

	\begin{equation} \label{ci}
	c_{in}(\alpha)=\frac{\int_{0}^{\alpha
	r(\mu_L)}r\,I(r)dr}{\int_{0}^{ r(\mu_L)}r\,I(r)dr} \end{equation}
	
	where $\mu_L$ is the detection threshold (set to 24.5
	mag$\cdot$arcsec$^{-2}$, mean value in our images) and
	$\alpha$ a parameter $0<\alpha<1$, appropriately chosen (it
	was set to 0.3, optimal value as described in Doi et
	al. \cite{Doi93}).\end{itemize}

	Mean surface brightnesses, radii (calculated as the semi-major
	axis of the isophote) and magnitudes inside the 24.5
	mag$\cdot$arcsec$^{-2}$ isophote, effective radii (see Paper
	I) and mean surface brightnesses inside the effective isophote
	have also been calculated.

	All these parameters are listed in Table \ref{tabres}. It was
	not possible to obtain reliable parameters for two objects,
	since their images were of very bad quality.

\subsection{Asymmetry parameter}	

	An asymmetry parameter $A$ was computed for each galaxy
	according to the definition established by Abraham et
	al. (\cite{Ab96a}). Each image was first smoothed with a
	Gaussian kernel of $\sigma$=1 pixel. After smoothing, it was
	rotated 180${\rm ^o}$ around the center of the object (this center
	was determined as the average of the inner isophotes of the
	galaxy). Finally, the rotated image was subtracted from the
	original. The parameter $A$ was calculated as:

	\begin{equation}
		A=\frac{\Sigma |(I_0-I_{180})|}{2\Sigma |I_0|}
	\end{equation}
	
	\noindent where the sum runs over all the pixels, I$_0$ is the
	intensity of the original smoothed image and I$_{180}$ the
	intensity of the rotated one. Since the absolute value of the
	flux of the self-subtracted image is used, the uncertainty in
	the sky value adds a positive $A$ signal. This effect was
	corrected calculating the parameter $A$ for a region of the
	sky of the same size of the galaxy aperture and then
	subtracting it from the one calculated for the galaxy. This
	coefficient is shown in Table \ref{tabres}.

\begin{table*}
\setcounter{table}{2}
\resizebox{\hsize}{!}{\includegraphics{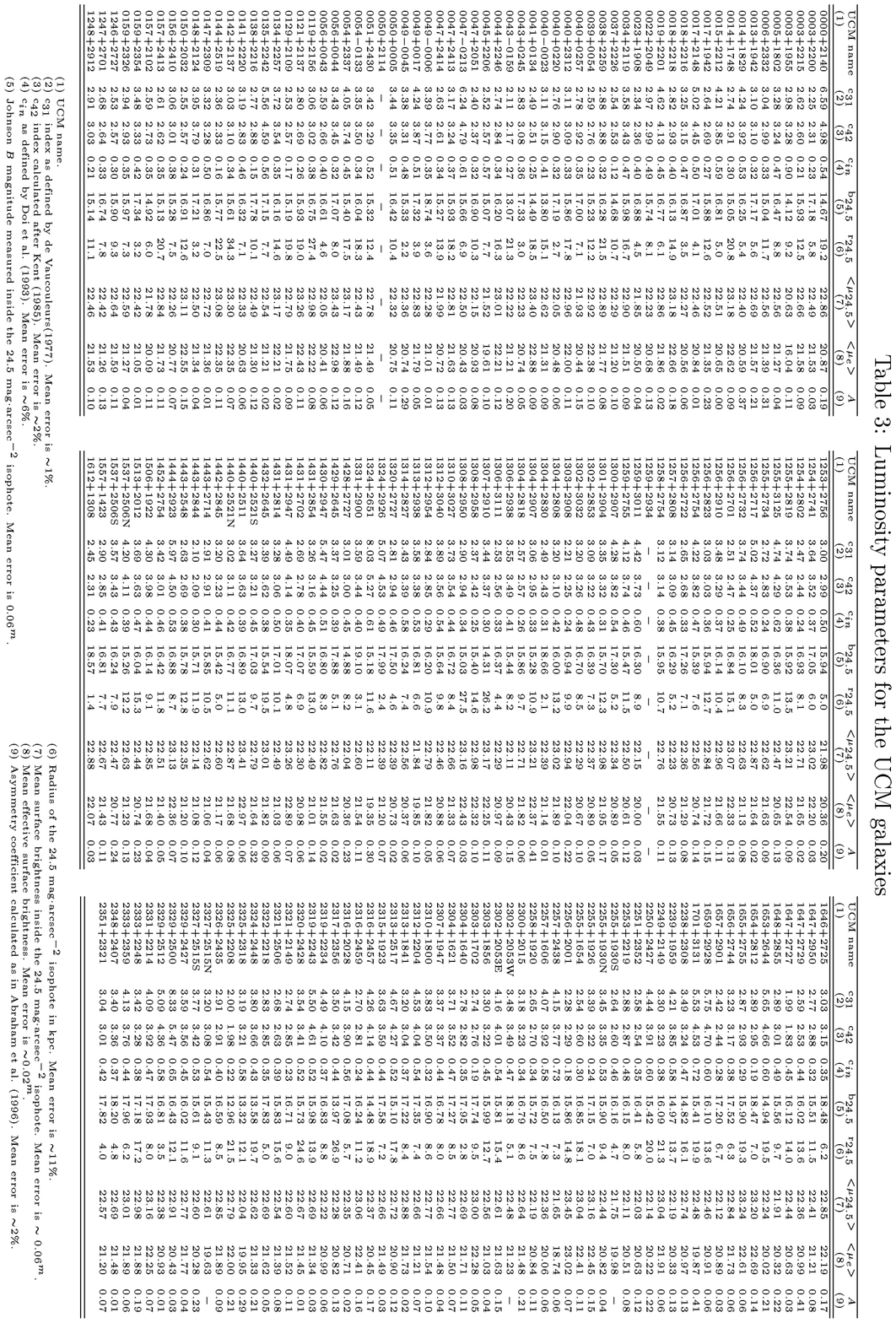}} 
\label{tabres} 
\end{table*}

\section{Result discussion}
\label{results}
\subsection{Bulge-disk parameters}
	
	Bulge-disk decomposition has been performed for a total number
	of 147 objects (77\% of the sample). The rest of the galaxy
	profiles were very distorted or the images did not present
	enough quality to attempt the fitting. None of the
	morphological types is segregated from this subsample.
	
\begin{figure}
\resizebox{\hsize}{!}{\psfig{file=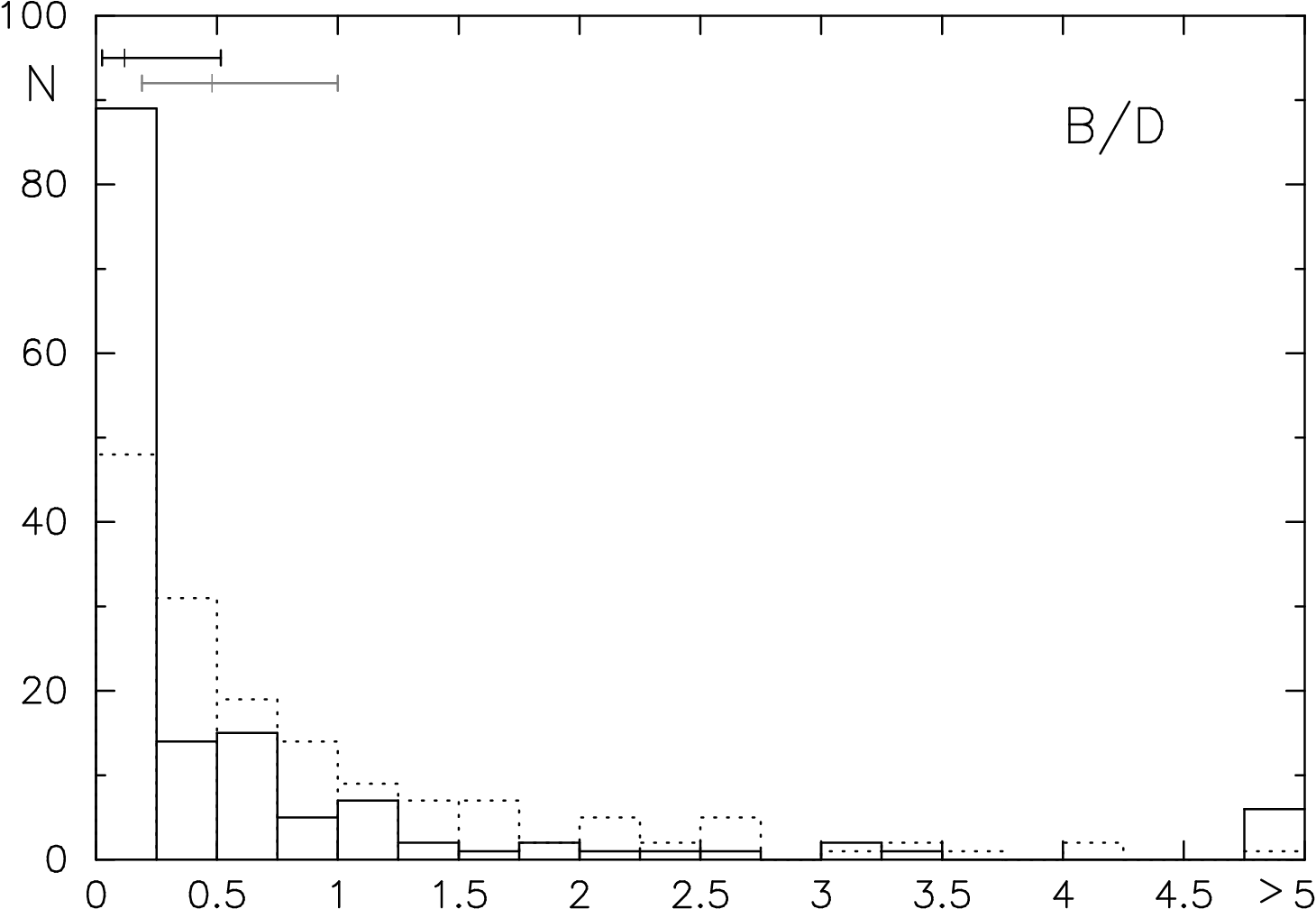,angle=-90}}
\caption{ 
Bulge-to-disk ratio histogram of the UCM Survey in the Johnson $B$
band. In this plot and hereafter, the median value and first quartiles
of the data will be shown at the top. Black lines correspond to $B$
band results; dotted lines for the histograms and grey lines for the
median and error bars will refer to the Gunn $r$ data from now on}
\label{bd}
\end{figure}

	Fig.~\ref{bd} shows the histogram of the bulge-to-disk
	luminosity ratio for the UCM Survey galaxies in the Johnson
	$B$ band; dotted lines in this picture and the next ones stand
	for the Gunn $r$ data (Vitores et al. \cite{Alv96a}). Median
	values (the upper corresponds to $B$ and the lower to $r$) and
	error bars referring to the first quartiles (black line for
	blue data and grey line for red results) are shown at the
	top. The mean B/D value is 0.40 with a standard deviation of
	0.65. This ratio is common for a Sb-Sbc galaxy, according to
	Kent (\cite{Ke85}). Special care should be taken with the B/D
	ratio when classifying galaxies, particularly when B/D$\ge$1.7
	(14 of our galaxies have a bulge-to-disk ratio above this
	value); based on this statement, this criterium has only been
	taken into account in galaxy profiles easily separable into
	clear bulge and disk components, where the concept of B/D
	ratio is meaningful (Simien \& de Vaucouleurs \cite{Si86},
	Schombert \& Bothun \cite{Sch87}).
	
	Overall, bulge-to-disk ratios based on $B$ images are lower
	than those calculated with the Gunn $r$ data. The difference
	could be, in part, due to the distinct methods used to fit the
	surface brightness profiles, being the seeing convolution
	treatment the main difference. We have performed a test on the
	artificial galaxies introduced in Sect~\ref{meth} not taking
	into account the seeing-dominated zone of the profiles; the
	bulge-to-disk ratios calculated in this case are, in average,
	$\sim$10\% larger than the values achieved using the seeing
	convolution. Therefore, the effect of seeing does not seem to
	cope with the whole difference between the $B$ and $r$
	bulge-to-disk ratios; on the contrary, it appears to be a real
	characteristic of the objects.

\begin{figure}
\resizebox{\hsize}{!}{\psfig{file=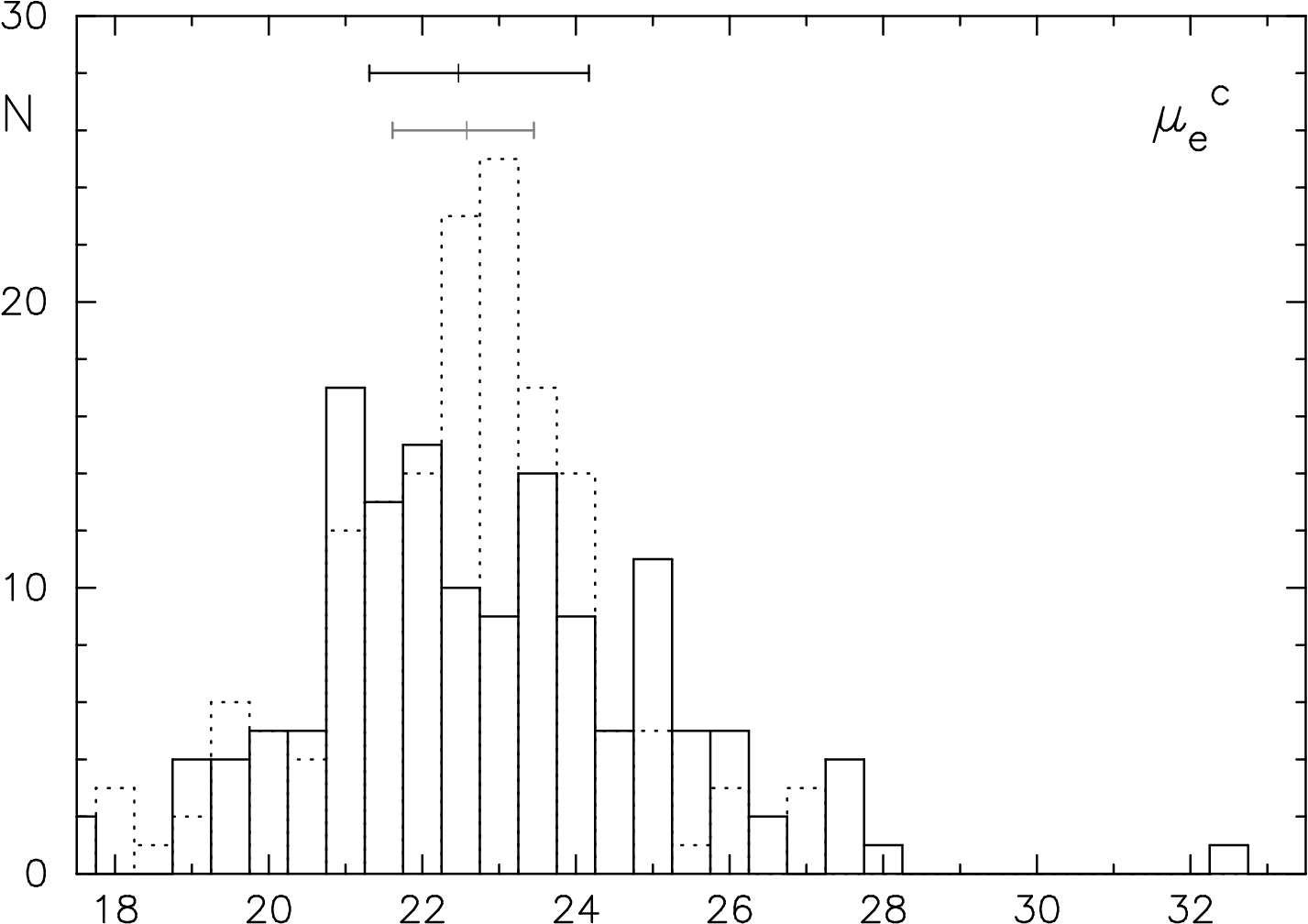,angle=-90}}
\caption{Histogram of the effective bulge surface brightness $\mu_e^c$ corrected 
for Galactic extinction}
\label{mu_e}
\end{figure}

\begin{figure}
\resizebox{\hsize}{!}{\psfig{file=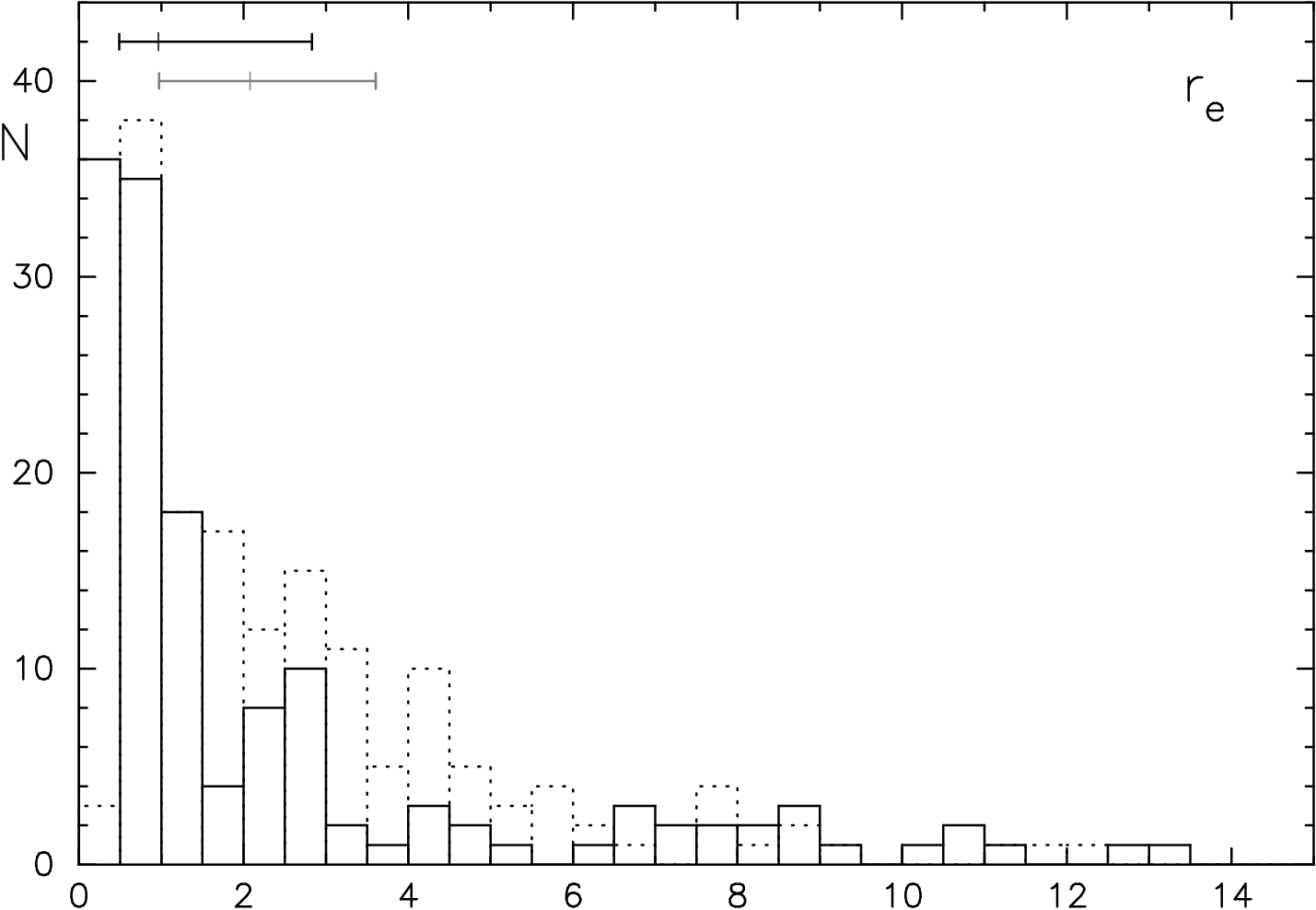,angle=-90}}
\caption{Histogram of the effective radius of the bulge component in kpc 
}
\label{r_e}
\end{figure}

\begin{figure}
\resizebox{\hsize}{!}{\psfig{file=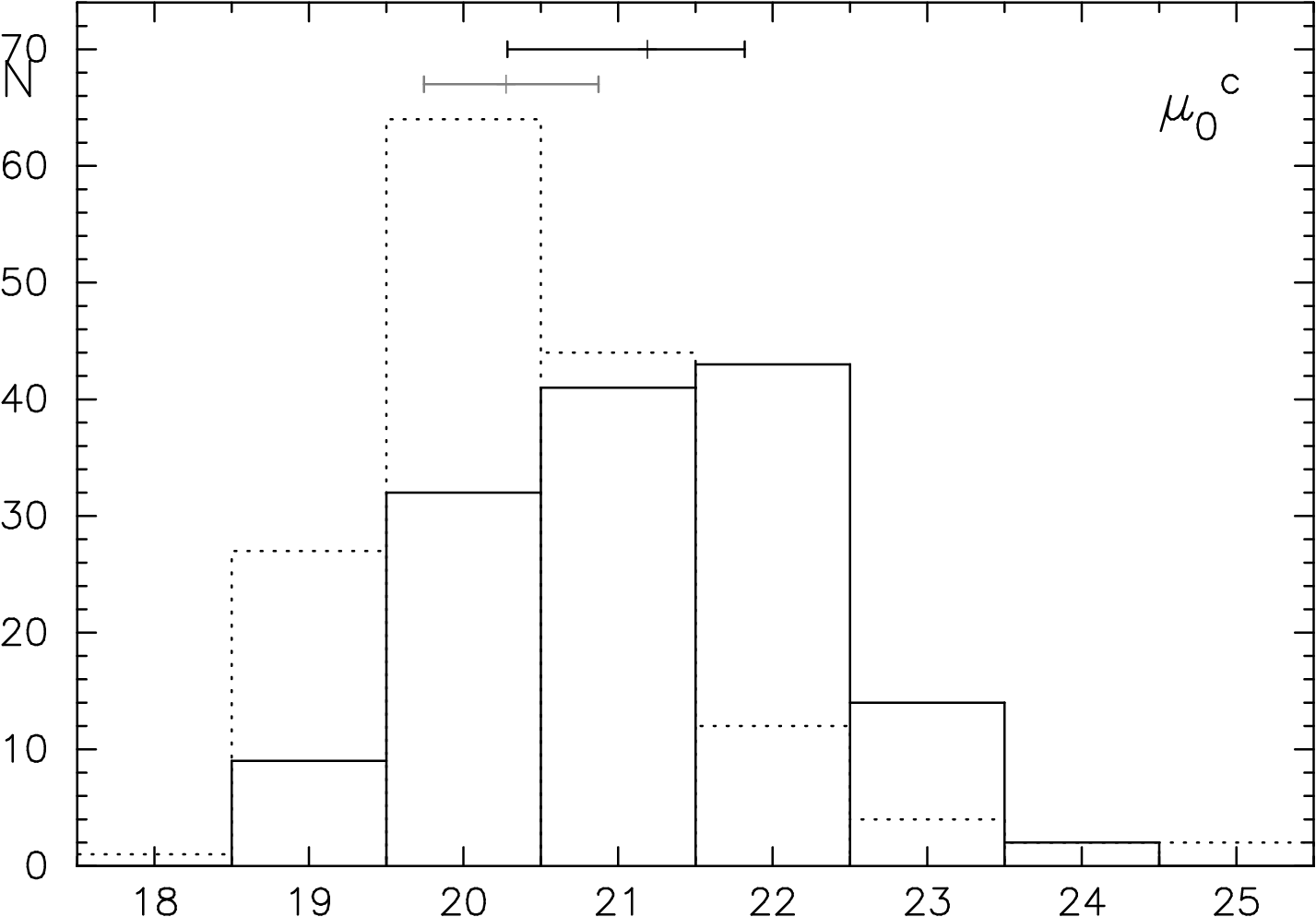,angle=-90}}
\caption{Histogram of the characteristic surface brightness of the disk $\mu_0^c$ corrected 
for Galactic extinction and inclination}
\label{mu_0}
\end{figure}

\begin{figure}
\resizebox{\hsize}{!}{\psfig{file=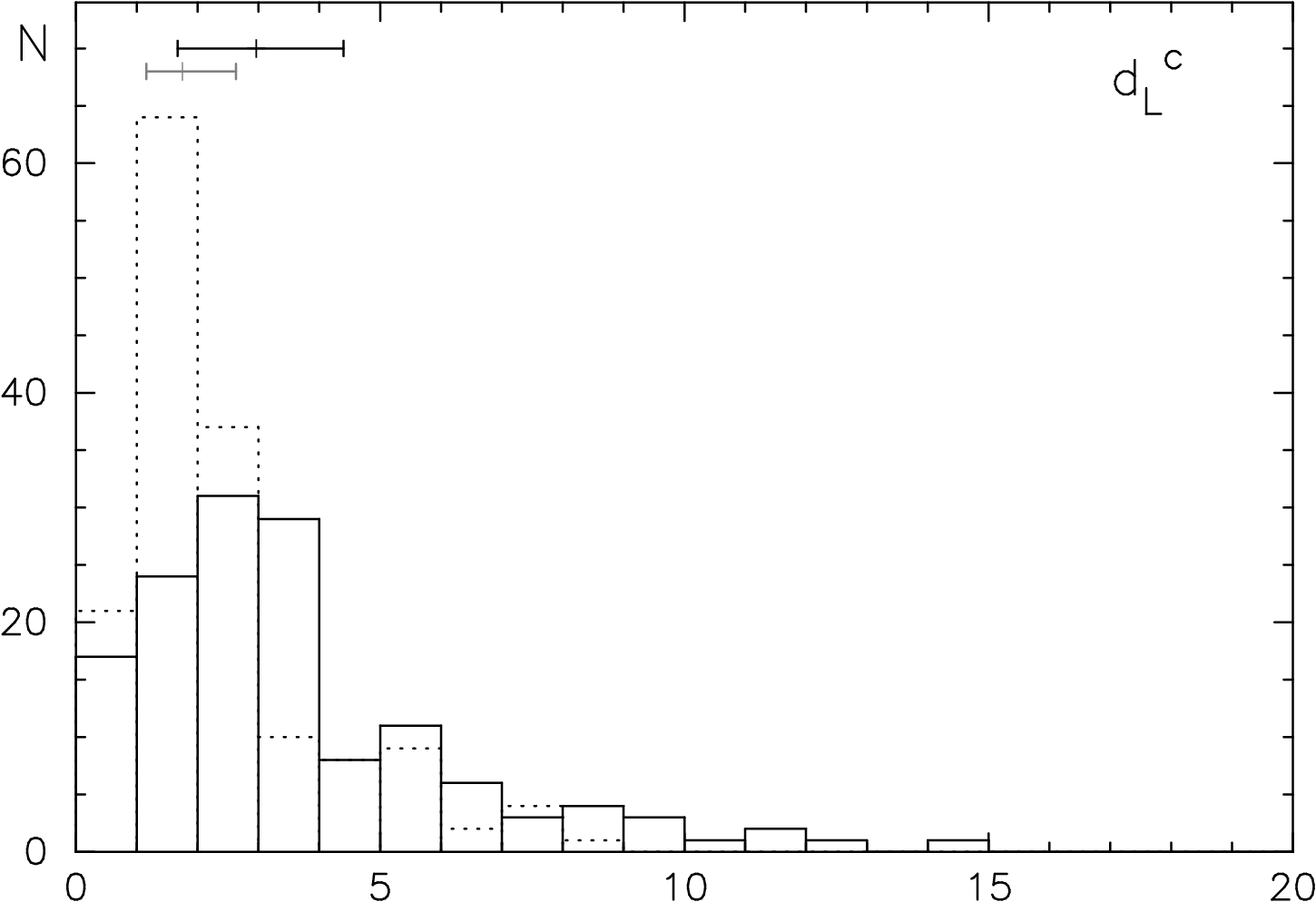,angle=-90}}
\caption{Histogram of the exponential scale length of the disk $d_L^c$ measured in kpc 
}
\label{d_L}
\end{figure}

	In Figs.~\ref{mu_e}, \ref{r_e}, \ref{mu_0} and \ref{d_L}, we
	show the histograms for the bulge and disk parameters
	$\mu_e^c$, $r_e$, $\mu_0^c$ and $d_L^c$, respectively. Scales
	are in kpc and surface brightnesses in
	mag$\cdot$arcsec$^{-2}$. Superindex $c$ denotes correction for
	Galactic extinction and inclination (in the disk typical
	surface brightness).

	The averaged $\mu_e^c$ value is 22.8$\pm$2.3
	mag$\cdot$arcsec$^{-2}$ (22.7$\pm$2.3 mag$\cdot$arcsec$^{-2}$
	if we only take into account the galaxies with B/D$<$1.7, the
	low B/D subsample hereafter), typical for a late-type spiral
	(Kent \cite{Ke85}; Simien \cite{Si89}). The typical scale of
	the bulge is in average 2.7 kpc (2.2 kpc for the low-B/D
	subsample), with a great dispersion ($\sigma$=4.8), but also
	common for a Sb-Sc galaxy. These values are very similar to
	the ones measured in the Gunn $r$ images, although there seems
	to be a lack of small bulges in the red data.

	The histogram of the characteristic surface brightness of the
	disk (Fig. \ref{mu_0}) is dominated by galaxies with
	$\mu_0^c$=21-22 mag$\cdot$arcsec$^{-2}$, with the average in
	21.1$\pm$1.1 mag$\cdot$arcsec$^{-2}$ (21.2$\pm$1.1
	mag$\cdot$arcsec$^{-2}$ for the low-B/D subsample). The narrow
	range of $\mu_0^c$ seems to support the existence of a
	universal central surface brightness for the disk, as proposed
	for normal spirals by Freeman (\cite{Fre70}) and confirmed by
	other authors (i.e., Boroson \cite{Bo81}, Simien \& de
	Vaucouleurs \cite{Si86}), although other works in the
	literature present samples of galaxies with a wider spread in
	$\mu_0^c$ (see, for example, McGaugh et al. \cite{McG95} or
	Beijersbergen et al. \cite{Bei99}). Our $\mu_0^c$ value is
	$\sim0.5^m$ brighter than the Freeman central surface
	brightness. Therefore, the UCM sample of star-forming galaxies
	appears to have brighter disks than those of normal spirals;
	this fact is probably related to the higher star-formation
	activity. Scale lengths are dominated by disks smaller than 4
	kpc (68\% of the total number of galaxies fitted), with mean
	3.6$\pm$2.6 kpc (the same for the low-B/D subsample). This
	value is higher than that found by Chitre et
	al. (\cite{Chi99}) for a sample of starburst galaxies in the
	Markarian sample ($d_L^c<$3 kpc), very similar to the averaged
	value found by Vennik et al. (\cite{Ven00}) for a sample of
	emission-line galaxies ($d_L^c$$\sim$2.7 kpc), although they
	only fit an exponential to the outer parts of the
	profiles. Our value is lower than the one found by de Jong
	(\cite{Jong96a}) for normal edge-on spirals ($d_L^c$$\sim$8
	kpc) -they argue that their selection biases against galaxies
	with low surface brightness and short scale lengths are
	large-. Other works (for example, Boroson \cite{Bo81}, Kent
	\cite{Ke85}, Bothun et al. \cite{Bot89}, Andredakis \& Sanders
	\cite{And94}) agree in placing our galaxies in the zone of
	short disk spirals, though one should be cautious against
	comparing scale lengths from different authors due to the
	subjective nature of disk parameters (Knapen \& van der Kruit
	\cite{Kna91} find discrepancies up to a factor of two in the
	scale lengths calculated from several authors).

	All of the above values are very similar to those found by
	Vitores et al. (\cite{Alv96b}) and place the UCM sample of
	galaxies in the zone of the late-type spirals, with small
	bulges and not very extended disks (Freeman
	\cite{Fre70}). Three remarks are interesting when comparing
	both sets of data. First, in the Gunn $r$ decomposition a
	lower bulge scale cut-off was observed (at $r_e$=0.5 kpc);
	this is not present in the $B$ band study. A possible
	explanation is the different handling performed with the
	seeing that allows the $B$ bulges to be smaller but brighter
	(seeing correction smoothes the profile; this was not the case
	with the Gunn $r$ bulge-disk decomposition, where the seeing
	effect was not taken into account directly), but this does not
	seem to cope with the whole difference. Second, both bands
	present a preference for disk scales around 2-3 kpc (larger
	disks in the blue band); very short disk scales and large ones
	are less frequent. Third, the difference between the surface
	brightness levels of the bulge and disk are of the order of
	the mean colour, around 0.5$^m$, as expected according to the
	averaged $B-r$ colour found in Fukugita et al. (\cite{Fuk95}).

\subsection{Geometric parameters}

\begin{figure}
\resizebox{\hsize}{!}{\psfig{file=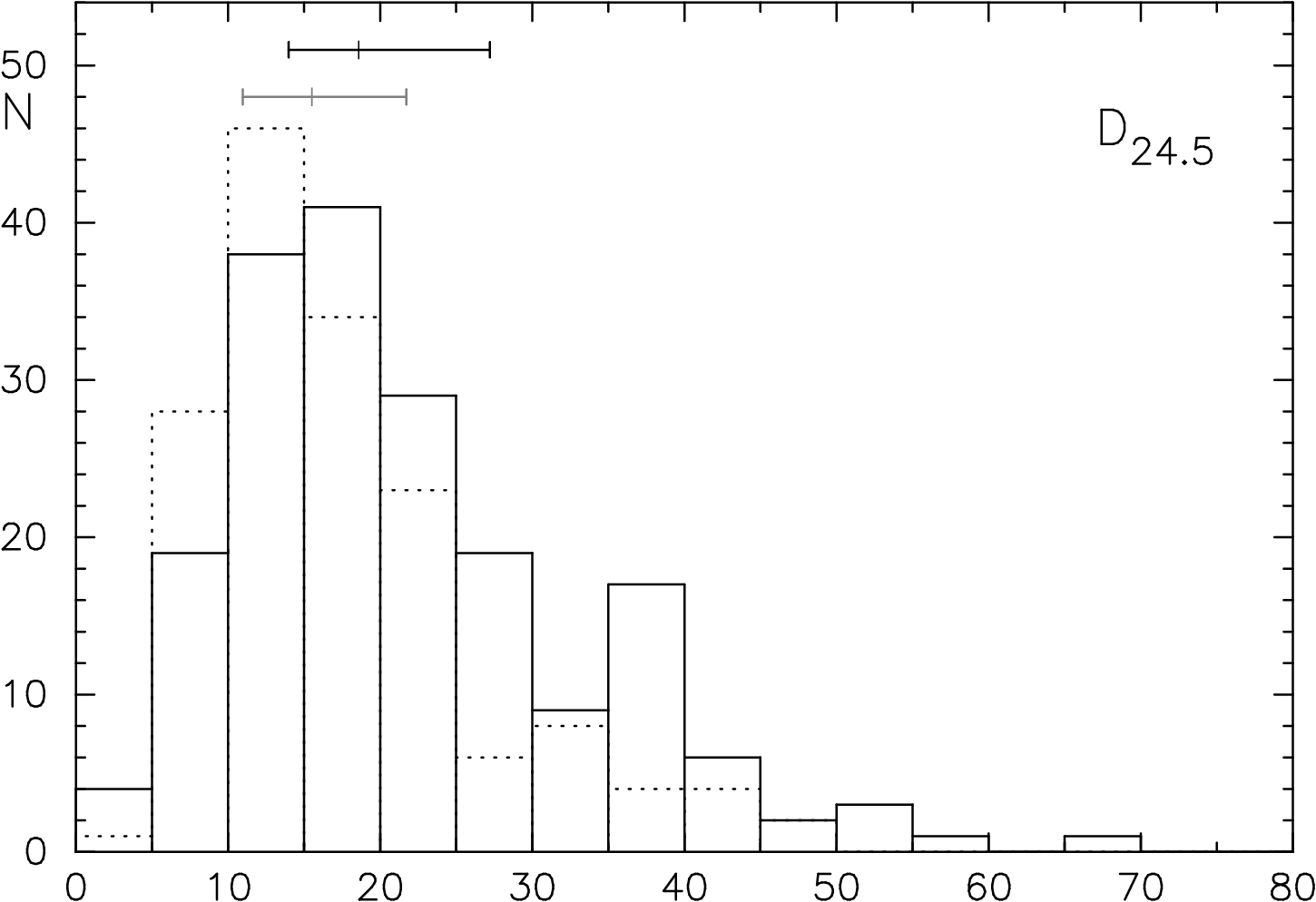,angle=-90}}
\caption{Histogram of the diameter of the 24.5 mag$\cdot$arcsec$^{-2}$ D$_{24.5}$
	isophote in kpc}
\label{diam}
\end{figure}

\begin{figure}
\resizebox{\hsize}{!}{\psfig{file=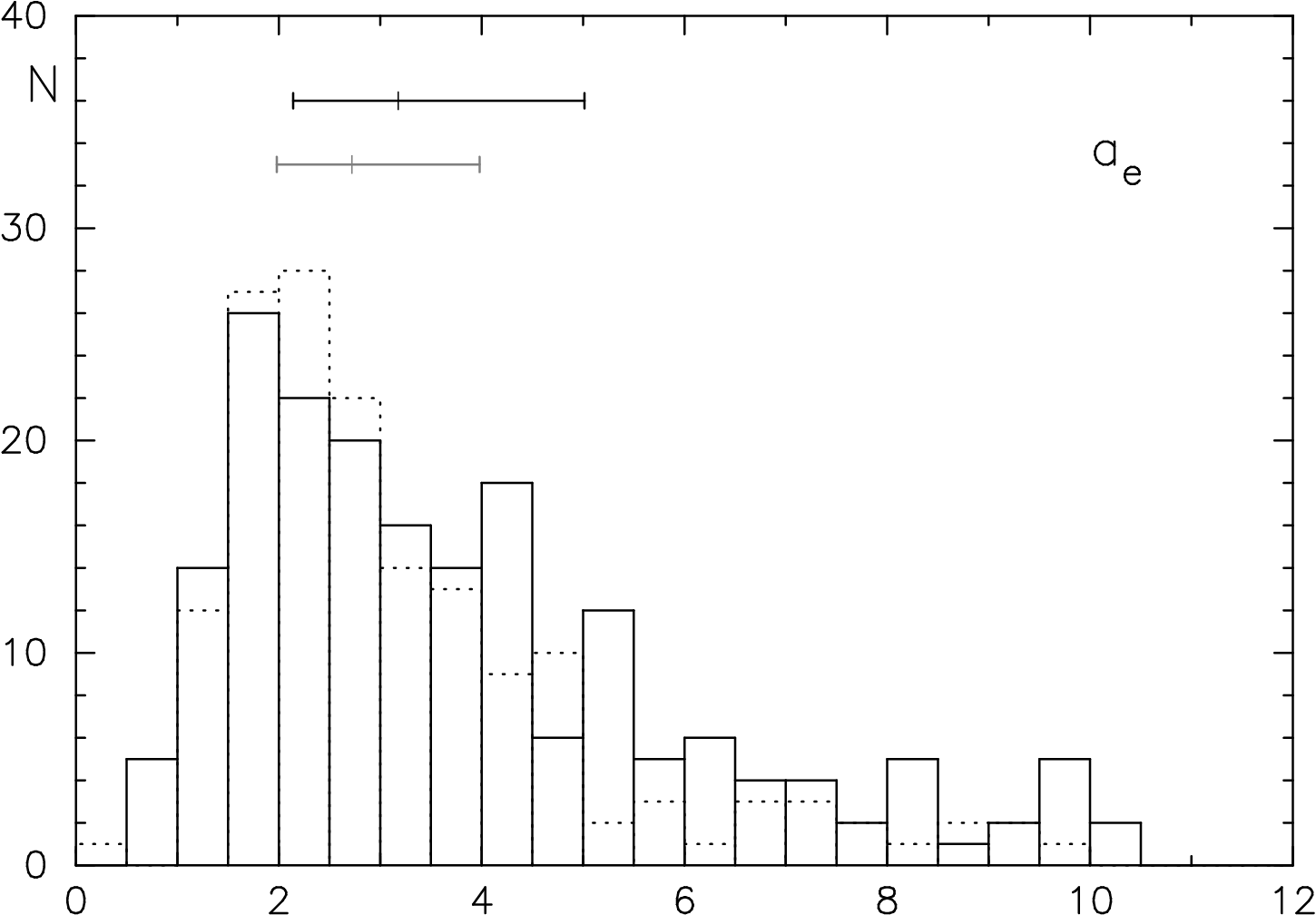,angle=-90}}
\caption{Histogram of effective radius $a_e$ in kpc
}
\label{radef}
\end{figure}

	In order to typify the size of the UCM galaxies, the
	histograms representing the diameter of the 24.5
	mag$\cdot$arcsec$^{-2}$ isophote D$_{24.5}$ and the effective
	radius $a_e$ (both in kpc) have been plotted in
	Figs.~\ref{diam} and \ref{radef}. The averaged diameter of the
	UCM objects is 22$\pm$12 kpc. Comparison with the red data has
	been established through the plot of the diameter of the 24
	mag$\cdot$arcsec$^{-2}$ Gunn $r$ isophote (that will be nearer
	to the 24.5 blue isophote than the corresponding red one,
	assuming a mean colour $B-r\sim0.5$). The mean effective
	radius $a_e$ is 3.8$\pm$2.3 kpc; this reflects the high degree
	of spatial luminosity concentration of our objects, most of
	them being starburst nuclei with a large emission arising from
	the center of the galaxy. Tentatively, UCM galaxies seem to be
	more extended in the blue band than in the red one (they show
	larger effective radius and diameters in $B$).

\begin{figure}
\resizebox{\hsize}{!}{\psfig{file=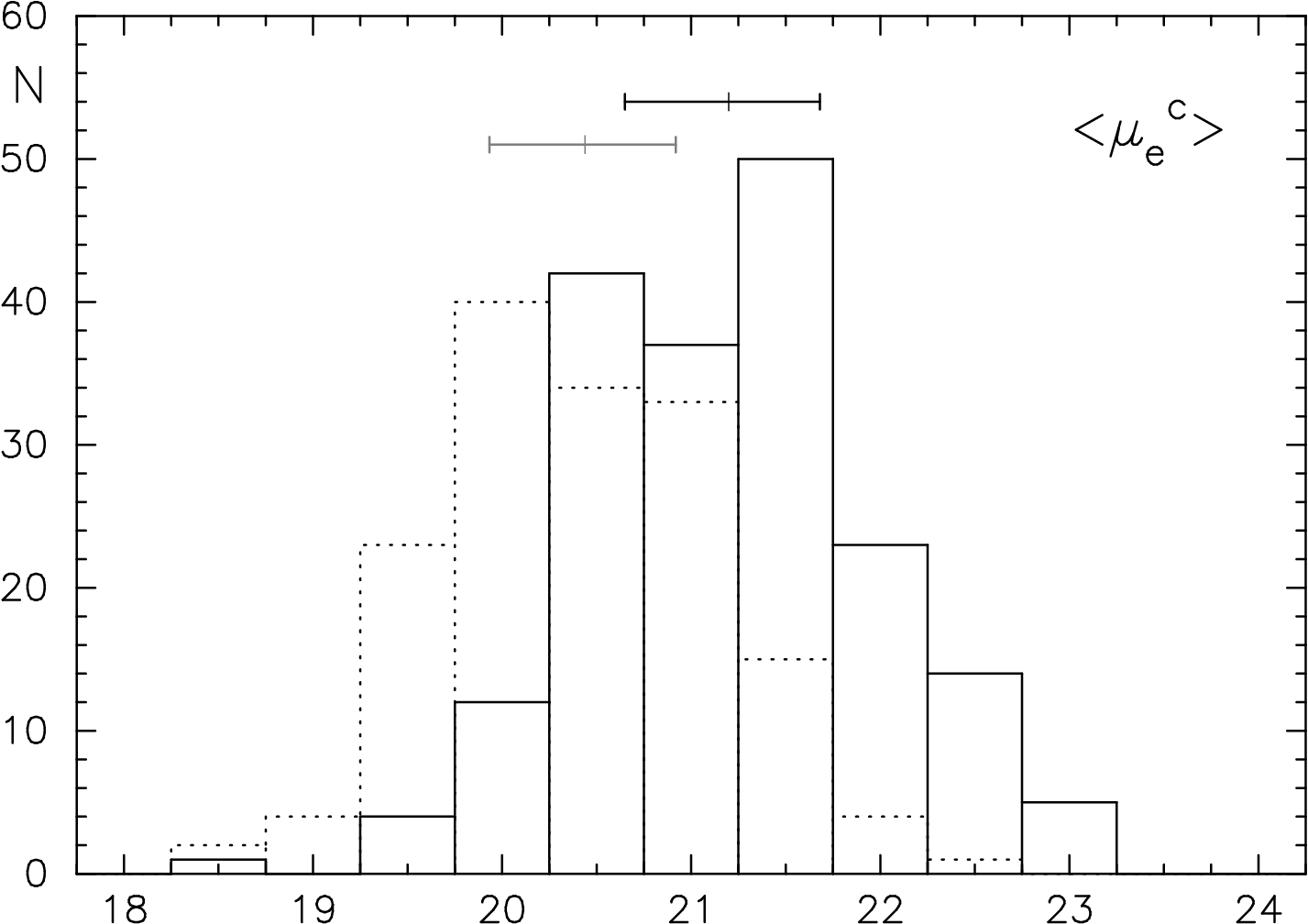,angle=-90}}
\caption{Histogram of the mean surface brightness inside the effective aperture $<\mu_e^c>$ corrected for Galactic extinction}
\label{muef}
\end{figure}

\begin{figure}
\resizebox{\hsize}{!}{\psfig{file=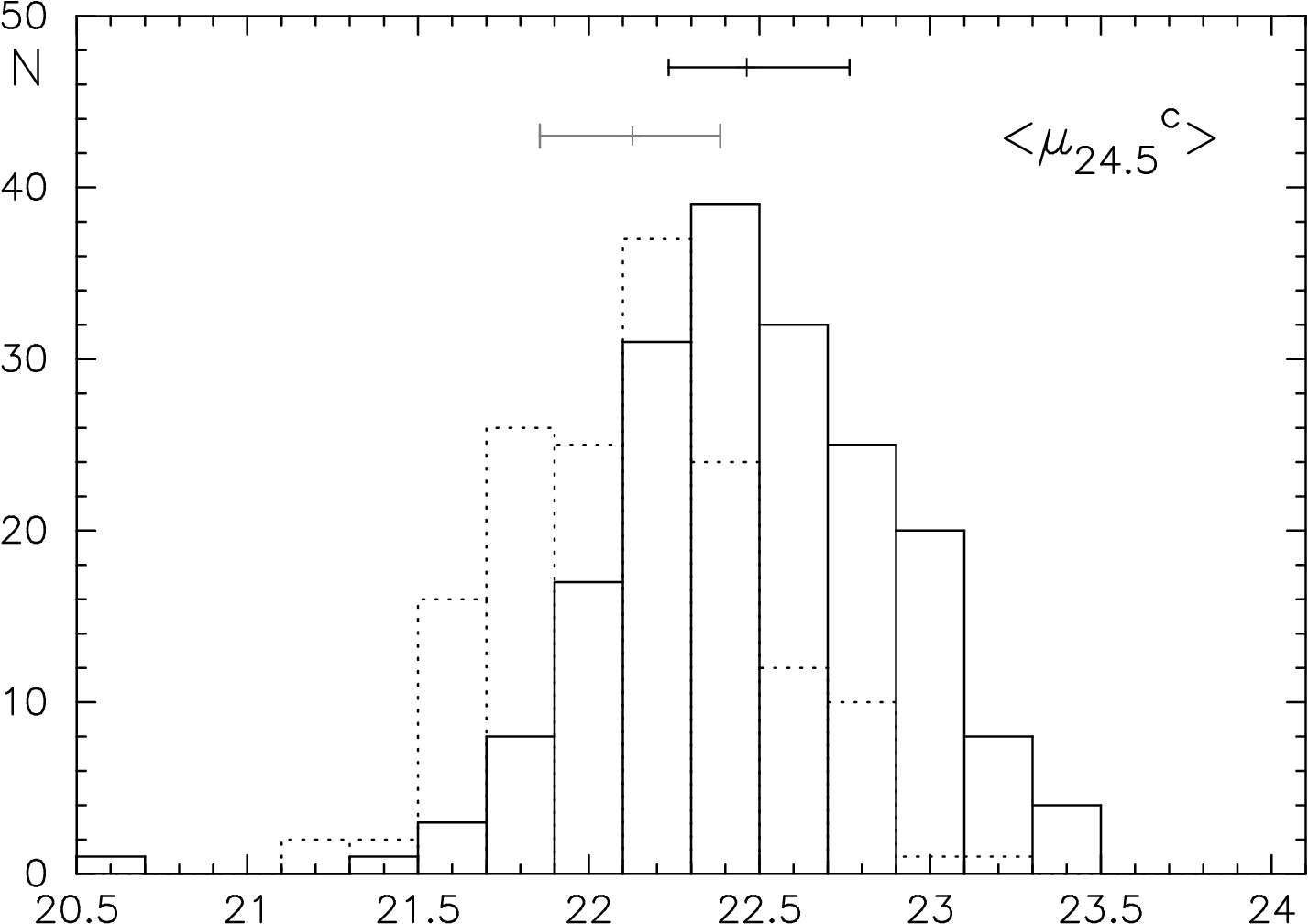,angle=-90}}
\caption{Histogram of the mean surface brightness inside the 24.5 isophote $<\mu_{24.5}^c>$ corrected for Galactic extinction}
\label{mu245}
\end{figure}

	Finally, we plot in Figs.~\ref{muef} and \ref{mu245} the mean
	effective and isophote 24.5 surface brightnesses in order to
	characterize the whole galaxy luminosity distribution. UCM
	objects show $<\mu_e^c>$=21.2$\pm$0.9 and
	$<\mu_{24.5}^c>$=22.5$\pm$0.4 (both in
	mag$\cdot$arcsec$^{-2}$), common value for normal galaxies
	(Doi et al. \cite{Doi93}). The difference between the Gunn $r$
	and the Johnson $B$ values ($\sim$0.5$^m$) is a common $B-r$
	colour for spirals (Fukugita et al. \cite{Fuk95}).

\subsection{Concentration indices and asymmetry coefficient}

\begin{figure}
\resizebox{\hsize}{!}{\psfig{file=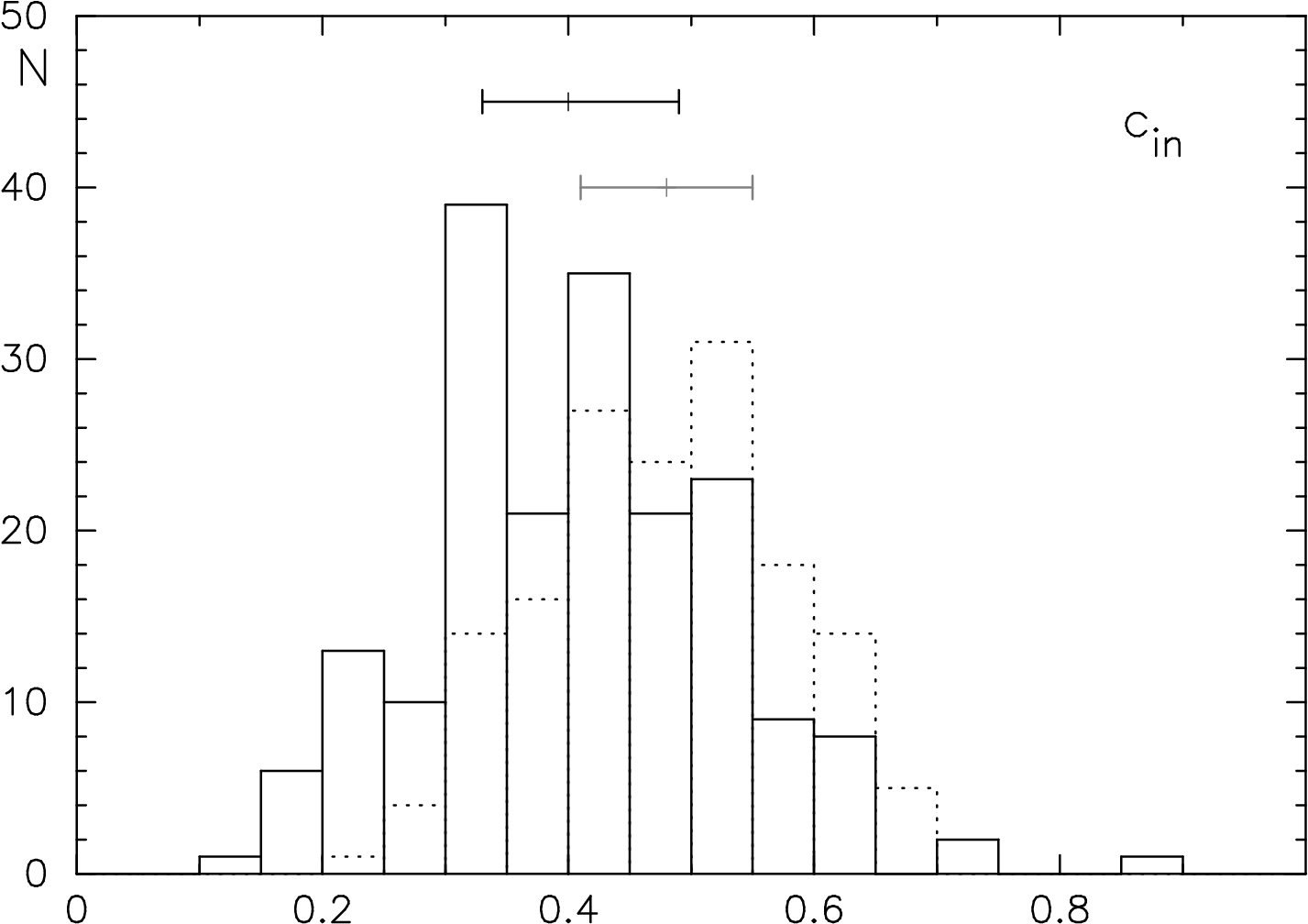,angle=-90}}
\caption{Histogram of the concentration index c$_{in}$ of the UCM sample}
\label{cin}
\end{figure}

\begin{figure}
\resizebox{\hsize}{!}{\psfig{file=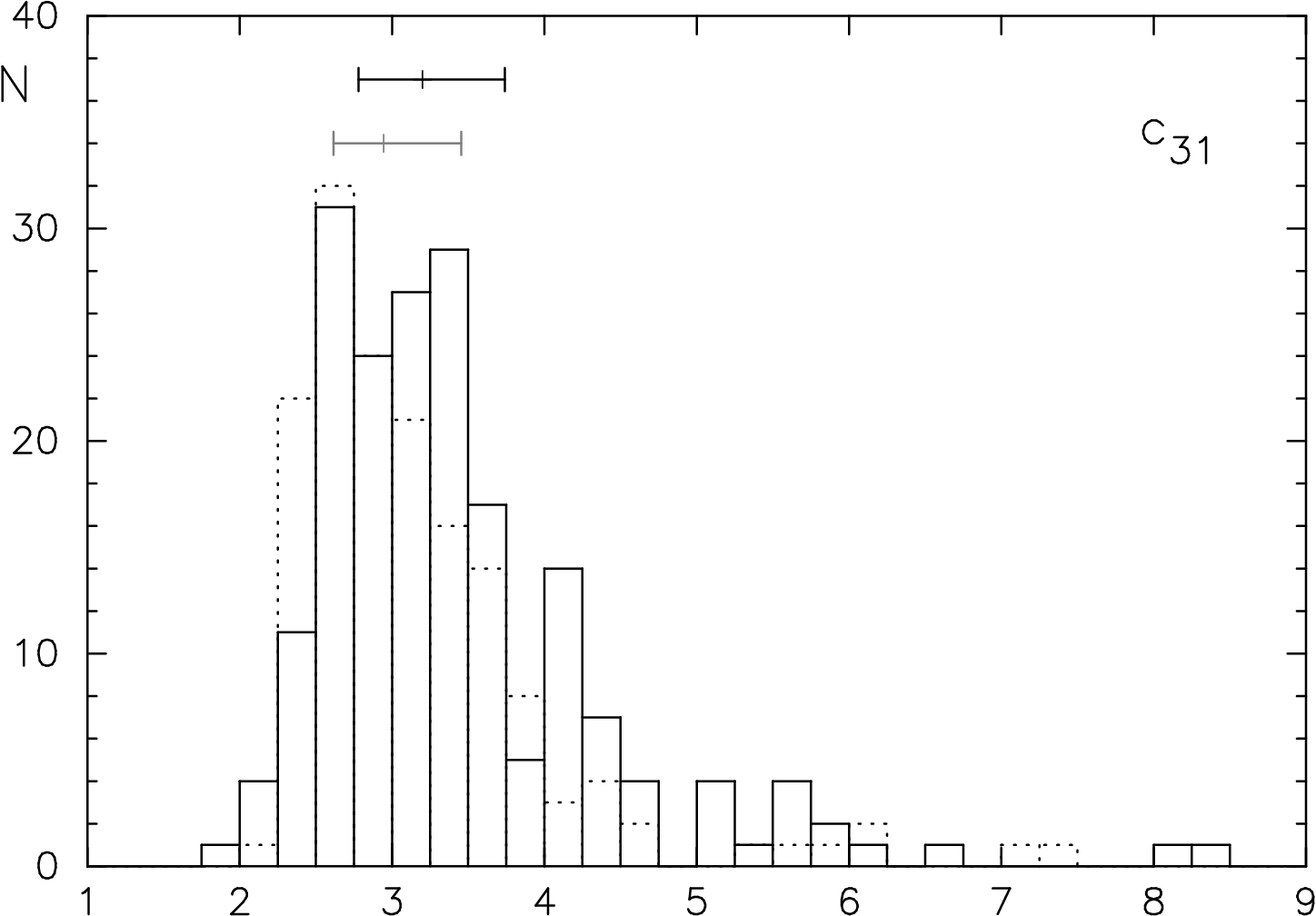,angle=-90}}
\caption{Histogram of the concentration index c$_{31}$ of the UCM sample}
\label{c31}
\end{figure}

\begin{figure}
\resizebox{\hsize}{!}{\psfig{file=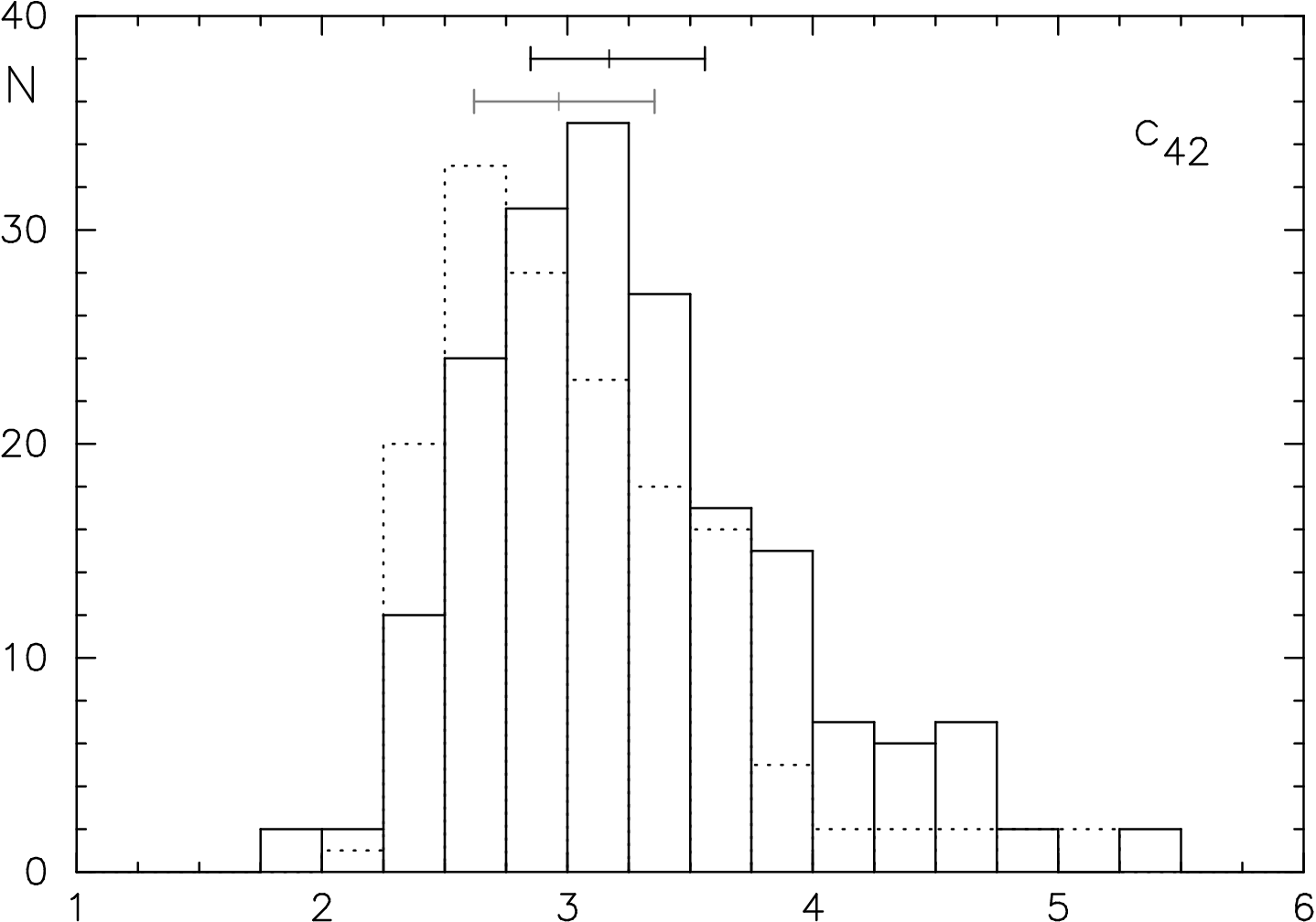,angle=-90}}
\caption{Histogram of the concentration index c$_{42}$ of the UCM sample}
\label{c42}
\end{figure}

	In the next 3 figures, labelled \ref{cin}, \ref{c31} and
	\ref{c42}, histograms of the concentration indices are
	shown. Mean values are c$_{in}$=0.41$\pm$0.12,
	c$_{31}$=3.4$\pm$1.0 and c$_{42}$=3.3$\pm$0.6. All of them are
	common values for spiral galaxies, corresponding approximately
	to a Hubble type of Sb (Doi et al. \cite{Doi93}, Gavazzi et
	al. \cite{Gav90}, Kent \cite{Ke85}, respectively for each
	concentration index). These values are higher than those
	measured in the Gunn $r$ images. The $B$ luminosity seems to
	be more concentrated in the inner parts than the $r$ one,
	although galaxies are more extended.

\begin{figure}
\resizebox{\hsize}{!}{\psfig{file=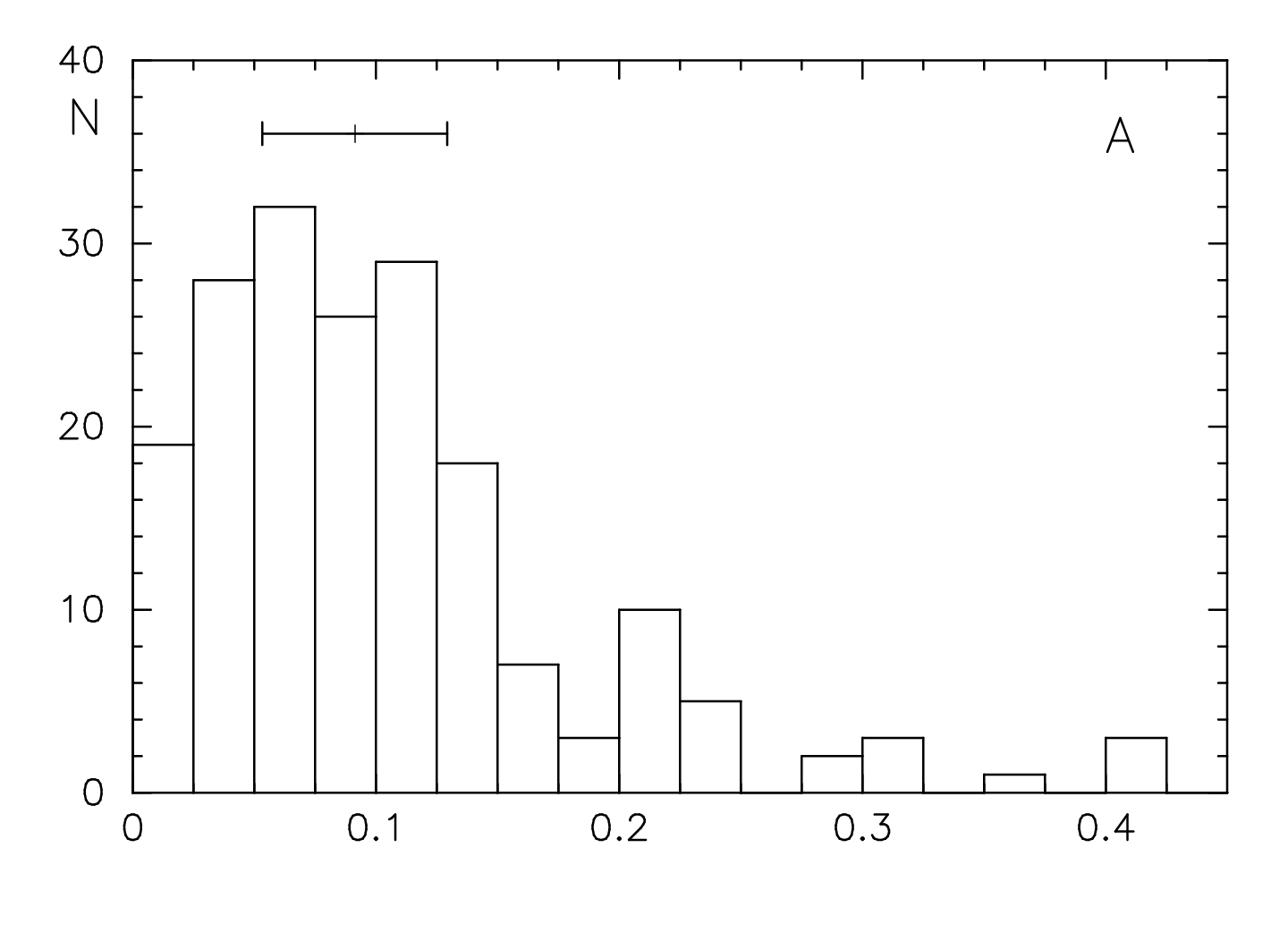,angle=-90}}
\caption{Histogram of the asymmetry coefficient $A$ of the UCM sample calculated after Abraham et al. \cite{Ab96a}}
\label{As}
\end{figure}

	Fig.~\ref{As} depicts the histogram of the asymmetry
	coefficient for the UCM sample.  The UCM sample is dominated
	by intermediately asymmetrical galaxies with mean
	0.10$\pm$0.08, lower than the value found by Bershady et
	al. (\cite{Ber00}) for a sample of normal local galaxies; this
	could be due to a difference in the calculation of $A$ or
	because their sample is composed by bright, large objects
	which probably have many asymmetrical features. This is what
	we should expect for spirals which have a certain axis
	symmetry although they present arms, bars or HII regions that
	enlarge the asymmetry coefficient. There is a lack of highly
	symmetrical objects, which correspond to elliptical galaxies,
	not present in our sample as it is composed by star-forming
	systems.

	All the previous results have been summarized in Table
	\ref{resumen} for a quick look, jointly with the Gunn $r$
	statistics.

\begin{table*}
\caption{Mean, median and standard deviation of the photometric parameters of the Johnson $B$ and Gunn $r$ (in brackets) images of the UCM Survey galaxies (scales are in kpc and surface brightnesses in mag$\cdot$arcsec$^{-2}$)}
\label{resumen}
\begin{tabular}{lcccc}
\hline
\vspace{-0.3cm} & & \\
Magnitudes & symbol & mean & st. dev. & median \\
\hline 
\hline 
{\bf Magnitudes}   &&&&\\
apparent magnitude & m$_B$  &   16.1 (15.5)   & 1.1 (1.0) &    16.1 (15.5)\\
absolute magnitude & M$_B$  &$-$19.9 ($-$20.5)& 1.1 (1.1) & $-$20.0 ($-$20.6)\\
\hline
{\bf B+D parameters}&&&&\\
bulge-to-disk ratio & B/D   & 0.40 (0.82) & 0.65 (0.98) & 0.12 (0.48)\\
effective bulge surface brightness & $\mu_e^c$ & 22.8 (22.6) & 2.3 (1.7) & 22.5 (22.6)\\
effective radius of the bulge      & $r_e$ & 2.7 (2.1) & 4.8 (3.3) & 1.0 (2.1) \\
disk face-on central surface brightness & $\mu_0^c$ & 21.1 (20.3) & 1.1 (1.1) & 21.2 (20.3) \\
exponential scale length of the disk & $d_L^c$ & 3.6 (1.8) & 2.6 (1.6) & 3.0 (1.8) \\
\hline
{\bf Geometric parameters} &&&&\\ diameter of the 24.5
mag$\cdot$arcsec$^{-2}$ isophote & D$_{24.5}$ & 22 (18) & 12 (9) & 19
(16) \\
\hline
{\bf Mean photometric parameters} &&&&\\
effective radius  & $a_e$ & 3.8 (3.3) & 2.3 (1.9) & 3.2 (2.7) \\
mean effective surface brightness & $<\mu_e^c>$& 21.2 (20.4) & 0.9 (0.7) & 21.2 (20.4) \\
isophote 24.5 mag$\cdot$arcsec$^{-2}$ mean surface brightness & $<\mu_{24.5}^c>$ & 22.5 (22.1) & 0.4 (0.4) & 22.5 (22.1) \\
\hline
{\bf Concentration indices}&&&&\\
concentration index ($\alpha=0.3$) & c$_{in}$  & 0.41 (0.48) & 0.12 (0.10) & 0.40 (0.48) \\
concentration index  & c$_{31}$& 3.4 (3.2) & 1.0 (0.9) & 3.2 (3.0) \\
concentration index & c$_{42}$ & 3.3 (3.1) & 0.6 (0.6) & 3.2 (3.0) \\
\hline
{\bf Asymmetry coefficient}&&&&\\
asymmetry coefficient       & $A$  & 0.10 (-) & 0.08 (-) & 0.09 (-) \\ 
\hline
\hline
\end{tabular}
\end{table*}

\subsection{Morphological classification}

  	A morphological classification of the UCM galaxies has been
  	carried out using 5 different criteria. These criteria were
  	already used by Vitores et al. (\cite{Alv96a}) with the Gunn
  	$r$ images, and are now applied to the Johnson $B$ data in
  	order to compare the results obtained with different
  	bandpasses. Besides, some galaxies not studied in Vitores et
  	al. (\cite{Alv96a}) have now been classified for the first
  	time (15\% of the sample). We outline the main features of the
  	classification criteria:
	
	\begin{itemize} \item the correlation between B/T ratio and
	Hubble type. B/T is defined as: \begin{equation}
	B/T=\frac{1}{(B/D)^{-1}+1} \end{equation} This correlation was
	studied by Kent (\cite{Ke85}, Fig. number 6) for a sample of
	bright galaxies in a red filter. It has been assumed that the
	behaviour of the correlation is very similar in the blue
	band.\item the dependence of the Hubble type on the position
	in the plane defined by the concentration index
	c$_{in}(\alpha)$ and $\mu_{24.5}^c$, first studied by Doi et
	al. (\cite{Doi93}). These authors argue that this criterium is
	rather insensitive to the colour band.\item the correlation
	between the concentration index c$_{31}$ and the Hubble type,
	as studied by Gavazzi et al. (\cite{Gav90}, Fig. 4b). \item
	the dependence of the morphological type on the concentration
	index c$_{42}$, established by Kent (\cite{Ke85}, Fig.
	11). \item the correlation between the mean surface brightness
	inside the effective isophote (corrected for Galactic
	extinction) and the Hubble type (Kent \cite{Ke85}, Fig.
	13). A mean correction of 0.5$^m$ due to the different
	bandpasses used in both works has been applied. \end{itemize}

	Visual inspection of each image was also used for the
	classification.

	In this work we utilized all these five criteria to classify
	the UCM galaxies in S0, Sa, Sb, Sc+ (Sc type or later) and Irr
	galaxies plus the BCD type (these galaxies were classified
	using spectroscopic confirmation available in Gallego et
	al. \cite{Gal96}); some galaxies were very distorted due to
	interactions and are marked in the result table as an
	independent class. The final Hubble type was established as
	that in which most criteria agree. This method is not
	completely objective and constitutes the main reason for the
	discrepancy between the classification using the Gunn $r$ data
	and that performed in this paper with the Johnson $B$ images.
	Table \ref{morph} presents the final classification in both
	bands. Fig.~\ref{5crit} shows the histograms and plots of the
	5 criteria used in the classification; in these plots the
	general trend of each parameter with the Hubble type can be
	seen, although great scatter and overlap between the different
	types are also present. Mean values will be shown in Table
	\ref{hubble}.

	Table \ref{numb} presents the number of UCM galaxies of each
	type in the Gunn $r$ and Johnson $B$ filters. A total number
	of 35 galaxies have been classified differently in the two
	bands, although the differences are always from one type to
	the contiguous (except in UCM2316+2028). Based on the Johnson
	$B$ data, 65\% of the whole sample is classified as Sb or
	later (61\% based on Gunn $r$ images). The percentage of
	barred galaxies is very similar in both bands (Johnson $B$
	9\%, Gunn $r$ 8\%); most of them are late-type spirals (47 \%
	are Sb galaxies and 35 \% Sc+). We have marked 6 clear
	interactions among the UCM galaxies (3\%), although there are
	more objects with tails or structures that could have been
	formed during an interaction. Seyfert 1 galaxies (6 objects)
	are all classified as S0, except one (UCM0003+1955) that is
	very bright and could not be classified; Sy 2 galaxies have
	been classified as Sa (1 object), Sb (3 objects) and Sc+ (3
	objects). These results are consistent with the ones found in
	the literature (see, for example, Hunt \& Malkan
	\cite{Hun99a}).

\vspace{-0.3cm}
\begin{table*}
\caption{Morphological classification of the UCM sample of galaxies}
\begin{tabular}{lcc|lcc|lcc}
\hline
{UCM name}& MpT($B$) & MpT ($r$) & {UCM name}& MpT($B$) & MpT ($r$) &  {UCM name}& MpT($B$) & MpT ($r$)   \\
 (1) & (2)  & (3) & (1) & (2) & (3) & (1) & (2) & (3) \\
\hline
\hline
0000+2140   &  INTER      &  ---    & 0141+2220   & Sa         &  Sb       &  1314+2827   & Sa         &  Sa    \\  
0003+2200   &  Sc+        &  Sc+    & 0142+2137   & SBb        &  SBb      &  1320+2727   & Sb         &  Sb    \\
0003+2215   &  Sc+        &  ---    & 0144+2519   & SBc+       &  SBc+(r)  &  1324+2926   & BCD        &  BCD   \\
0003+1955   &  ---        &  ---    & 0147+2309   & Sa         &  Sa       &  1324+2651   & INTER      &  ---   \\
0005+1802   &  Sb         &  ---    & 0148+2124   & BCD        &  BCD      &  1331+2900   & BCD        &  BCD   \\
0006+2332   &  Sb         &  ---    & 0150+2032   & Sc+        &  Sc+      &  1428+2727   & Irr        &  Sc+   \\
0013+1942   &  Sc+        &  Sc+    & 0156+2410   & Sb         &  Sc+      &  1429+2645   & Sb         &  Sc+   \\
0014+1829   &  Sa         &  Sa     & 0157+2413   & Sc+        &  Sc+      &  1430+2947   & S0         &  S0    \\
0014+1748   &  SBb        &  SBb    & 0157+2102   & Sb         &  Sb       &  1431+2854   & Sb         &  Sb    \\
0015+2212   &  Sa         &  Sa     & 0159+2354   & Sb         &  Sa       &  1431+2702   & Sa         &  Sb    \\
0017+1942   &  Sc+        &  Sc+    & 0159+2326   & Sc+        &  Sc+      &  1431+2947   & BCD        &  BCD   \\
0017+2148   &  Sa         &  ---    & 1246+2727   & Irr        &  ---      &  1431+2814   & Sb         &  Sa    \\
0018+2216   &  Sb         &  Sb     & 1247+2701   & Sc+        &  Sc+      &  1432+2645   & SBb        &  SBb   \\
0018+2218   &  Sb         &  ---    & 1248+2912   & SBb        &  ---      &  1440+2521S  & Sb         &  Sb    \\
0019+2201   &  Sb         &  Sc+    & 1253+2756   & Sa         &  Sa       &  1440+2511   & Sb         &  Sb    \\
0022+2049   &  Sb         &  Sb     & 1254+2741   & Sb         &  Sb       &  1440+2521N  & Sb         &  Sa    \\
0023+1908   &  Sc+        &  ---    & 1254+2802   & Sc+        &  Sc+      &  1442+2845   & Sb         &  Sb    \\
0034+2119   &  SBc+       &  ---    & 1255+2819   & Sb         &  Sb       &  1443+2714   & Sa         &  Sa    \\
0037+2226   &  SBc+       &  ---    & 1255+3125   & Sa         &  Sa       &  1443+2844   & SBc+       &  SBc+  \\
0038+2259   &  Sb         &  Sa     & 1255+2734   & Sc+        &  Irr      &  1443+2548   & Sc+        &  Sc+   \\
0039+0054   &  Sc+        &  ---    & 1256+2717   & S0         &  ---      &  1444+2923   & S0         &  S0    \\
0040+0257   &  Sb         &  Sc+    & 1256+2732   & INTER      &  ---      &  1452+2754   & Sb         &  Sb    \\
0040+2312   &  Sc+        &  ---    & 1256+2701   & Sc+        &  Irr      &  1506+1922   & Sb         &  Sb    \\
0040+0220   &  Sc+        &  Sb     & 1256+2910   & Sb         &  Sb       &  1513+2012   & Sa         &  S0    \\
0040$-$0023 &  Sc+        &  ---    & 1256+2823   & Sb         &  Sb       &  1537+2506N  & SBb        &  SBb   \\
0041+0134   &  Sc+        &  ---    & 1256+2754   & Sa         &  Sa       &  1537+2506S  & SBa        &  SBa   \\
0043+0245   &  Sc+        &  ---    & 1256+2722   & Sc+        &  Sc+      &  1557+1423   & Sb         &  Sb    \\
0043$-$0159 &  Sc+        &  ---    & 1257+2808   & Sb         &  Sa       &  1612+1308   & BCD        &  BCD   \\
0044+2246   &  Sb         &  Sb     & 1258+2754   & Sb         &  Sb       &  1646+2725   & Sc+        &  Sc+   \\
0045+2206   &  INTER      &         & 1259+2934   & Sb         &  Sb       &  1647+2950   & Sc+        &  Sc+   \\
0047+2051   &  Sc+        &  Sc+    & 1259+3011   & Sa         &  Sa       &  1647+2729   & Sb        &  Sb     \\
0047$-$0213 &  S0         &  Sa     & 1259+2755   & Sa         &  Sa       &  1647+2727   & Sb        &  Sa     \\
0047+2413   &  Sa         &  Sa     & 1300+2907   & Sa         &  Sb       &  1648+2855   & Sa        &  Sa     \\
0047+2414   &  Sc+        &  ---    & 1301+2904   & Sb         &  Sb       &  1653+2644   & INTER     &  ---    \\
0049$-$0006 &  BCD        &  BCD    & 1302+2853   & Sb         &  Sa       &  1654+2812   & Sc+       &  Sc+    \\
0049+0017   &  Sb         &  Sc+    & 1302+3032   & Sa         &  ---      &  1655+2755   & Sc+       &  Sb     \\
0049$-$0045 &  Sb         &  ---    & 1303+2908   & Irr        &  Irr      &  1656+2744   & Sa        &  Sa     \\
0050+0005   &  Sa         &  Sa     & 1304+2808   & Sb         &  Sa       &  1657+2901   & Sb        &  Sc+    \\
0050+2114   &  Sa         &  Sa     & 1304+2830   & BCD        &  BCD      &  1659+2928   & SB0       &  SB0    \\
0051+2430   &  Sa         &  ---    & 1304+2907   & Irr        &  Irr      &  1701+3131   & S0        &  S0     \\
0054$-$0133 &  Sb         &  ---    & 1304+2818   & Sc+        &  Sc+      &  2238+2308   & Sa(r)     &  Sa     \\
0054+2337   &  Sc+        &  ---    & 1306+2938   & SBb        &  Sb       &  2239+1959   & S0        &  S0     \\
0056+0044   &  Irr        &  Irr    & 1306+3111   & Sc+        &  Sc+      &  2249+2149   & Sb        &  Sa     \\
0056+0043   &  Sb         &  Sc+    & 1307+2910   & SBb        &  SBb      &  2250+2427   & Sa        &  Sa     \\
0119+2156   &  Sb         &  Sc+    & 1308+2958   & Sc+        &  Sc+      &  2251+2352   & Sc+       &  Sc+    \\
0121+2137   &  Sc+        &  Sc+    & 1308+2950   & SBb        &  SBb      &  2253+2219   & Sa        &  Sa     \\
0129+2109   &  SBc+       &  ---    & 1310+3027   & Sb         &  Sa       &  2255+1930S  & Sb        &  Sb     \\
0134+2257   &  Sb         &  ---    & 1312+3040   & Sa         &  Sa       &  2255+1930N  & Sb        &  Sb     \\
0135+2242   &  S0         &  S0     & 1312+2954   & Sc+        &  Sc+      &  2255+1926   & Sb        &  Sc+    \\
0138+2216   &  Sc+        &  ---    & 1313+2938   & Sa         &  Sa       &  2255+1654   & Sc+       &  Sc+    \\
\hline                                                                               
\hline                                                                               
\end{tabular}
\end{table*}                                                                         
\begin{table*}  
\setcounter{table}{4}
\caption{continued}                                          
\begin{tabular}{lcc|lcc|lcc}
\hline
{UCM name}& MpT($B$) & MpT ($r$) & {UCM name}& MpT($B$) & MpT ($r$) &  {UCM name}& MpT($B$) & MpT ($r$)   \\
 (1) & (2)  & (3) & (1) & (2) & (3) & (1) & (2) & (3) \\                                                               
\hline
\hline
2256+2001   & Sc+      &   Sc+     & 2313+1841   & Sb       &   Sb   &  2325+2318   & INTER    &   ---    \\ 
2257+2438   & S0       &   S0      & 2313+2517   & Sa       &   ---  &  2325+2208   & SBc+     &   SBc+   \\ 
2257+1606   & S0       &   ---     & 2315+1923   & Sb       &   Sa   &  2326+2435   & Sb       &   Sa     \\ 
2258+1920   & Sc+      &   Sc+     & 2316+2457   & SBa      &   SBa  &  2327+2515N  & Sb       &   Sb     \\ 
2300+2015   & Sb       &   Sb      & 2316+2459   & Sc+      &   Sc+  &  2327+2515S  & S0       &   S0     \\ 
2302+2053W  & Sb       &   Sb      & 2316+2028   & Sa       &   Sc+  &  2329+2427   & Sb       &   Sb     \\ 
2302+2053E  & Sb       &   Sb      & 2317+2356   & Sa       &   Sa   &  2329+2500   & S0(r)    &   S0(r)  \\ 
2303+1856   & Sa       &   Sa      & 2319+2234   & Sb       &   Sc+  &  2329+2512   & Sa       &   Sa     \\ 
2303+1702   & Sc+      &   Sc+     & 2319+2243   & S0       &   S0   &  2331+2214   & Sb       &   Sb     \\ 
2304+1640   & BCD      &   BCD     & 2320+2428   & Sa       &   Sa   &  2333+2248   & Sc+      &   Sc+    \\ 
2304+1621   & Sa       &   Sa      & 2321+2149   & Sc+      &   Sc+  &  2333+2359   & S0a      &   S0     \\ 
2307+1947   & Sb       &   Sb      & 2321+2506   & Sc+      &   Sc+  &  2348+2407   & Sa       &   Sa     \\ 
2310+1800   & Sb       &   Sc+     & 2322+2218   & Sc+      &   Sc+  &  2351+2321   & Sb       &   Sb     \\ 
2312+2204   & Sa       &   ---     & 2324+2448   & Sb       &   Sc+  &              &          &          \\ 
\hline 														  
\hline
\label{morph}
\end{tabular}\\
\vspace{-0.5cm}
\setcounter{table}{4}
\caption{(1) UCM name. (2) Morphological type established using 5 different criteria based on luminosity concentration and bulge-disk decomposition applied to the Johnson $B$ images. (3)  Morphological type established using 5 different criteria based on luminosity concentration and bulge-disk decomposition applied to the Gunn $r$ images}
\end{table*}

\begin{table}
\caption{Hubble types for the UCM galaxies}
\label{numb}
\begin{tabular}{ccccccccc}
\hline
\vspace{-0.3cm} & & \\
Filter & S0 & Sa & Sb & Sc+ & Irr & BCD & Int & Total\\
\hline 
\hline 
$B$ & 14    & 38   & 69     & 50     &  5    &  8    &  6    & 190 \\
            & 7\% & 20\% & 36\% & 26\% & 3\% & 4\% & 3\% &  \\
$r$ & 12    & 41     & 43     & 46     &  5    &  8    &  $-$  & 155 \\ 
            & 7\% & 27\% & 28\% & 30\% & 3\% & 5\% &     & \\
\hline
\hline
\end{tabular}
\setcounter{table}{5}
\caption{Number of galaxies and percentage of the UCM galaxies according to their Hubble type in the Johnson $B$ and the Gunn $r$ bandpasses}
\end{table}

\section{Correlations between parameters}
\label{correlations}
	
	We plot in Figs.~\ref{cor1} and \ref{cor2b} the relationships
	between absolute $B$ magnitude and the size of the galaxy
	(24.5 isophote diameter) and also between M$_B$ and the
	distribution of light (concentration index
	c$_{31}$). Information about morphological classification is
	also shown.

	There is a tight correlation between M$_B$ and D$_{24.5}$. A
	least-square fit to our data leads to:

	\begin{equation}
	log\,D_{24.5}=(-2.65\pm0.16)+(-0.20\pm0.01)\,M_B
	 \end{equation}
	
	The slope is very similar to the value expected for a constant
	luminosity-area ratio ($log\,D=C-0.2\,M$). Therefore, despite
	there is a great variety in morphological and spectroscopic
	types, a uniformity in the mean surface brightness is
	exhibited, as was also proved with the red data (Vitores et
	al.  \cite{Alv96b} found a slope value of
	$-$0.21$\pm$0.01). The fit gives a mean surface brightness
	value of $-$13.3 mag$\cdot$kpc$^{-2}$.

	In Fig.~\ref{cor2b} a general trend between the concentration
	index c$_{31}$, the absolute $B$ magnitude M$_B$ and the
	morphological type is apparent. Early-type galaxies show
	medium-high magnitudes and high concentration indices. If we
	move downwards to the zone of low concentration index we find
	spirals, from Sa to late-type. Finally, BCDs have c$_{31}$
	values typical for spirals but are fainter than normal
	galaxies.

	Fig.~\ref{cor2c} shows the segregation in morphological type
	in a c$_{in}$ versus A diagram. In this plot and the next,
	median values for the different morphological types are
	plotted with a black dot; ellipse semi-axes are the $\sigma$
	of each parameter. There is a clear trend from left to right
	in decreasing Hubble type.  S0 galaxies are placed in the high
	symmetry-high c$_{in}$ zone. BCDs also appear as highly
	symmetrical objects. On the other hand, irregulars are shown
	as highly asymmetrical objects in the top-left zone of the
	plot and interactive systems are located among the most
	asymmetrical galaxies. A trend can be also remarked in the
	spiral sequence: early-type galaxies are more symmetrical than
	late-type ones (due to the presence of more HII regions, for
	example).

\setcounter{figure}{14}
\begin{figure}
\resizebox{\hsize}{!}{\psfig{file=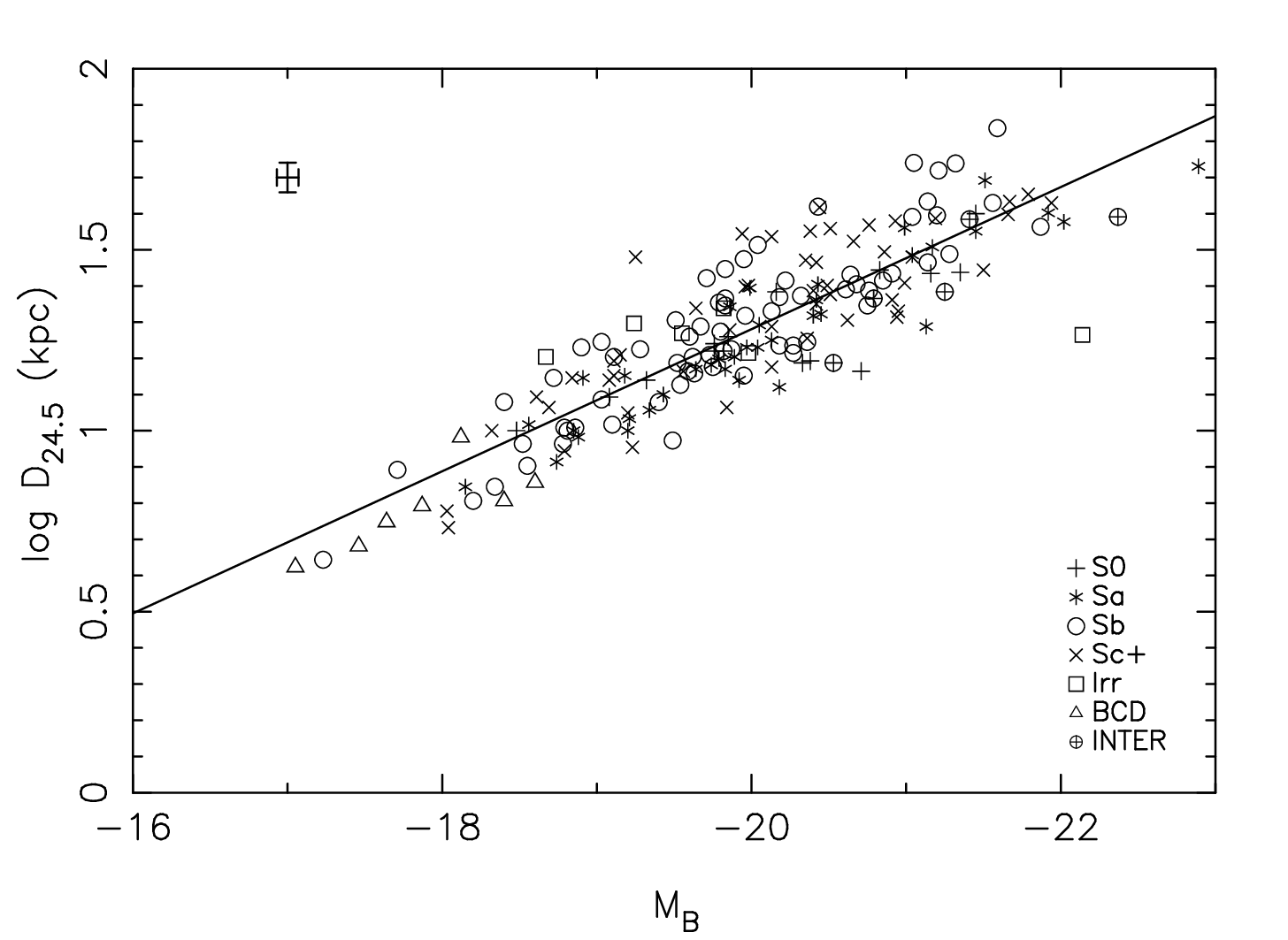,angle=-90}}
\caption{Relationship between the size of the UCM galaxies represented by the diameter of the 24.5 isophote D$_{24.5}$ and the total $B$ luminosity of the object M$_B$. A least-square fit to the data is also plotted}
\label{cor1}
\end{figure}

\begin{figure}
\resizebox{\hsize}{!}{\psfig{file=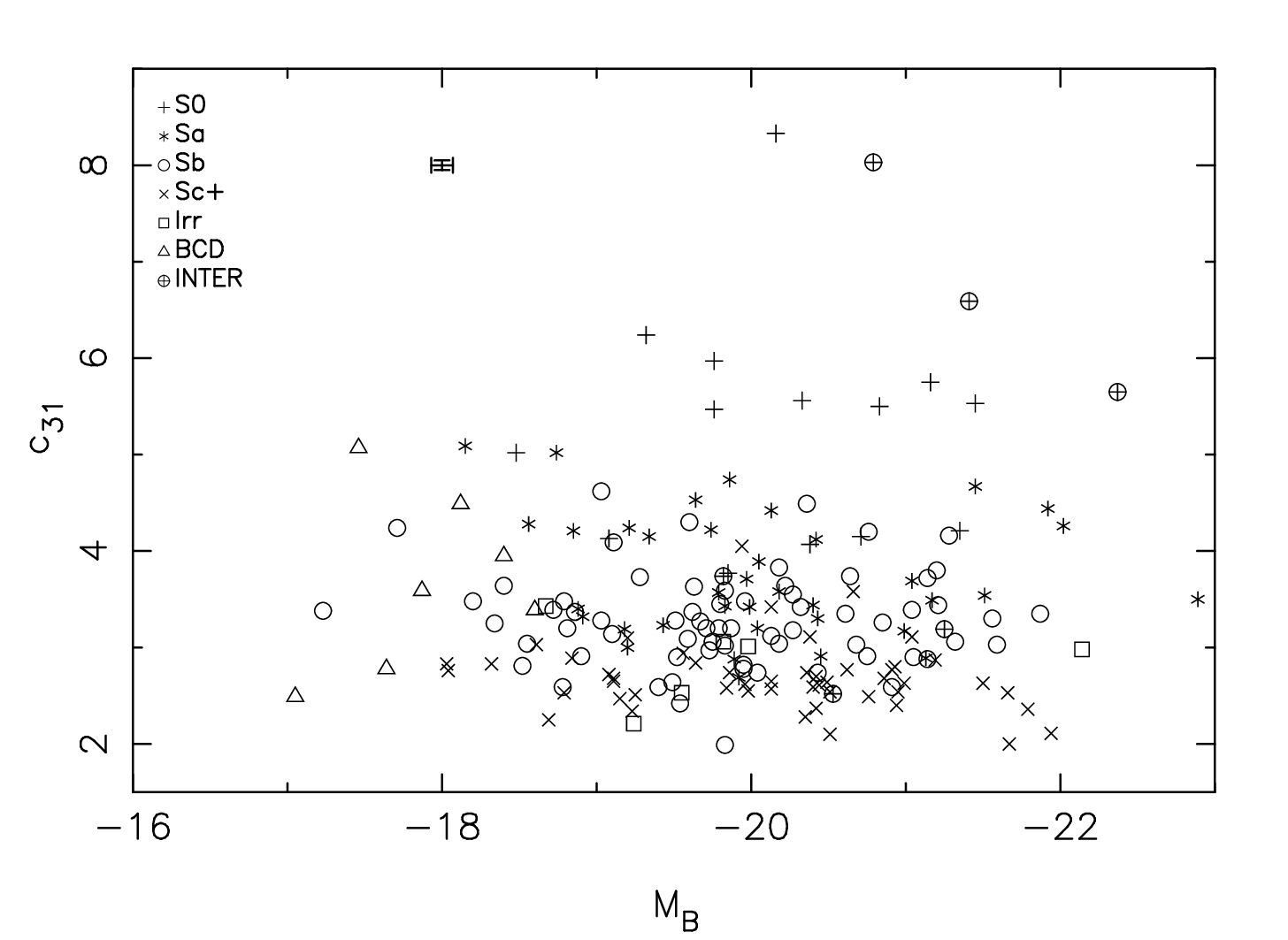,angle=-90}}
\caption{Concentration index c$_{31}$ versus absolute magnitude M$_B$. Different symbols stand for distinct morphological types}
\label{cor2b}
\end{figure}

\begin{figure}
\resizebox{\hsize}{!}{\psfig{file=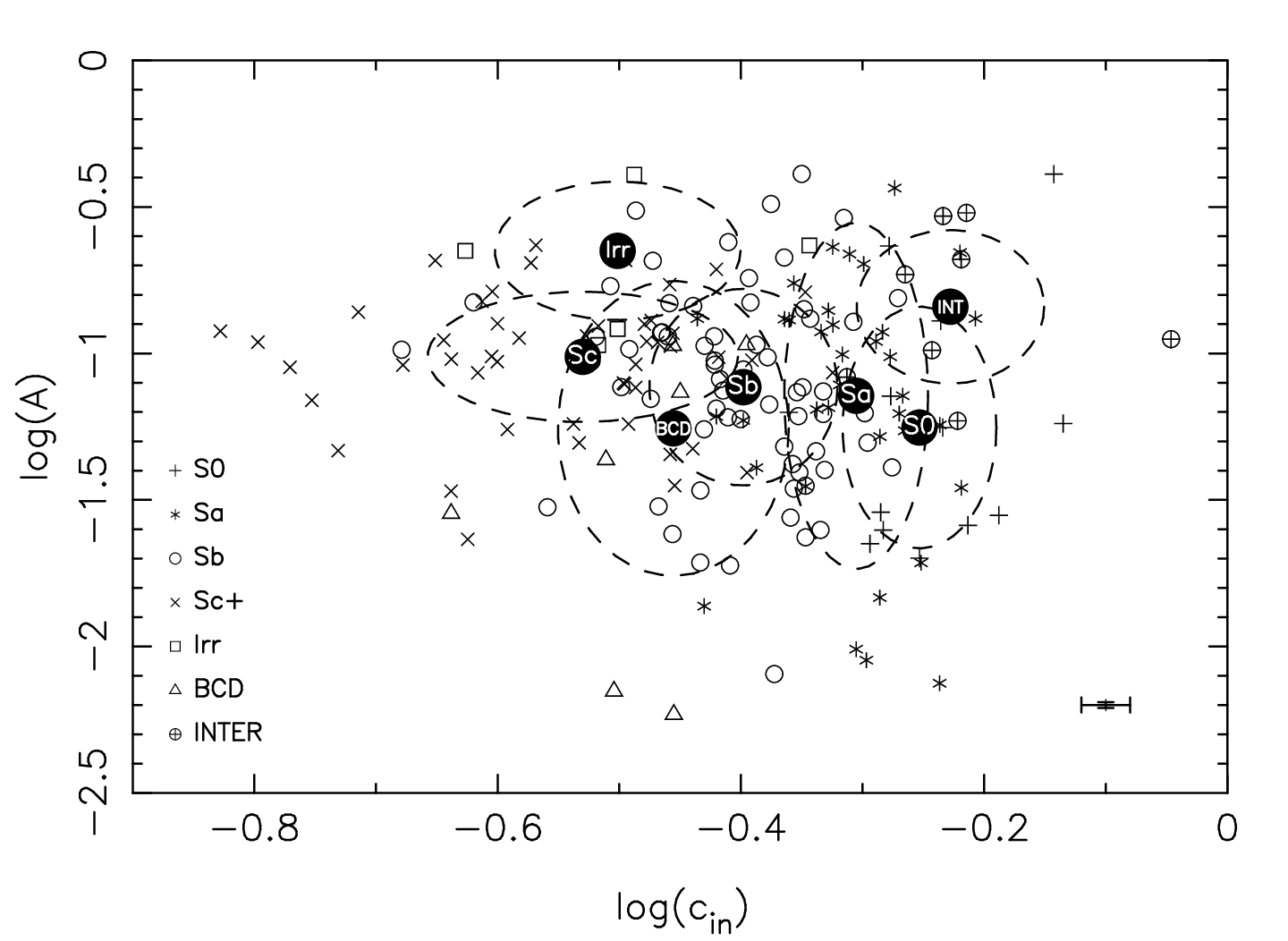,angle=-90}}
\caption{Concentration index c$_{in}$ versus asymmetry coefficient. Different symbols stand for distinct morphological types. Ellipses show the median and standard deviation of each parameter for the different morphological types}
\label{cor2c}
\end{figure}

	Fig. \ref{5crit} showed that there is a clear correlation
	between the concentration indices and Hubble type. This trend
	is also observed with the asymmetry coefficient. Table
	\ref{hubble} presents the mean values of the bulge-to-disk
	ratio, mean effective surface brightness, concentration
	indices and asymmetry coefficient of each Hubble type. The
	statistics of $A$ have been split into barred and
	non-barred objects; barred galaxies are more asymmetrical than
	non-barred ones.

\begin{table}
\caption{Averaged values of each Hubble type}
\label{hubble}
\begin{tabular}{cccccccc}
\hline
\vspace{-0.3cm} & & \\
Parameter & S0  & Sa   & Sb   & Sc+  & Irr  & BCD  & Int \\
\hline 
\hline
B/T      & 0.53 & 0.39 & 0.16 & 0.04 & 0.07 & 0.41 & 0.30\\
\hline 
$\mu_e$  & 20.6 & 20.7 & 21.3 & 21.5 & 21.8 & 21.5 & 20.1\\ 
\hline 
c$_{31}$ & 5.3  & 3.8  & 3.3  & 2.7  & 2.9  & 3.5  & 5.0 \\
\hline 
c$_{42}$ & 4.3  & 3.5  & 3.2  & 2.7  & 2.8  & 3.4  & 4.0 \\
\hline 
c$_{in}$ & 0.58 & 0.50 & 0.40 & 0.29 & 0.33 & 0.35 & 0.57\\
\hline  
A        & 0.09 & 0.10 & 0.10 & 0.11 & 0.21 & 0.06 & 0.20\\
(barred) & 0.06 & 0.15 & 0.11 & 0.12 &  $-$ & $-$  & $-$  \\
\hline
\hline
\end{tabular}
\setcounter{table}{6}
\end{table}

\begin{figure}
\resizebox{\hsize}{!}{\psfig{file=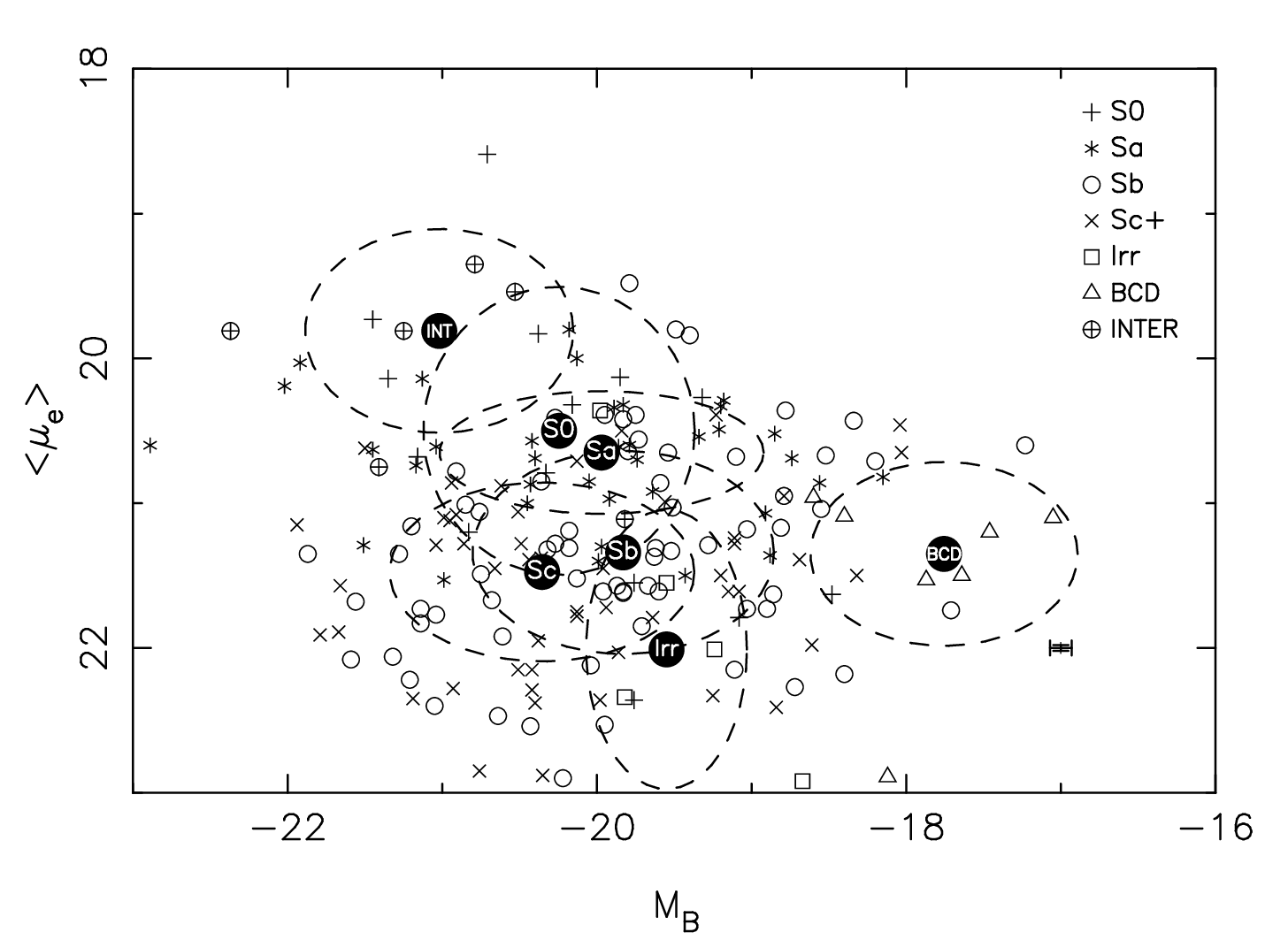,angle=-90}}
\caption{Mean effective surface brightness versus absolute magnitude. Ellipses show the median and standard deviation of each parameter for the different morphological types}
\label{bon}
\end{figure}

	Fig.~\ref{bon} depicts the absolute $B$ magnitude of the UCM
	objects versus the mean effective surface
	brightness. Early-type galaxies appear as bright, high surface
	brightness objects. Late-type spirals have lower $\mu_e$,
	although no significant difference in M$_B$ is present. BCDs
	are clearly segregated due to their faintness. Irregulars and
	interactive systems show also a distinctive surface
	brightness.

\section{Summary and conclusions}

	We have carried out a morphological study of the UCM Survey
	galaxies based on Johnson $B$ imaging. This paper, jointly
	with Paper I (P\'erez-Gonz\'alez et al. \cite{yor}), have
	analyzed the main features of the UCM sample concerning
	integrated and surface photometry in the $B$ bandpass.

	Paper I presented integrated apparent and absolute $B$
	luminosities as well as isophote 24 mag$\cdot$arcsec$^{-2}$
	magnitudes. Effective radii and $B-r$ colours were also
	calculated in this first release. In the present paper we have
	outlined the main results concerning bulge-disk decomposition
	values, ellipticities, position angles, concentration indices,
	mean photometric radii and surface brightnesses and asymmetry
	coefficients.

	All the above information has been used to perform the
	morphological classification of the UCM galaxies. The sample
	is dominated by late Hubble type objects (65\% being Sb or
	later). We have not found a great difference between this
	classification and the results achieved with the Gunn $r$ data
	(Vitores et al. \cite{Alv96a}, \cite{Alv96b}).
	
	Our galaxies are characterized by shorter disks than those of
	normal spirals. Besides, they seem to be objects with a high
	luminosity concentration. A preliminary comparison between the
	characteristics of the sample in the red and blue bandpasses,
	yields to the result that emission-line galaxies have a higher
	concentration of blue light than red light in the inner parts
	of the objects; on the contrary, they also seem to be more
	extended in $B$ than in $r$.

	Finally, a size-luminosity correlation has been outlined. We
	have also presented several plots where a morphological
	segregation is patent. These plots involve information about
	luminosity, concentration of light and asymmetry.

	In next papers we will face a most exhaustive comparison
	between bands, including colour gradients and a stellar
	population study. Likewise, we will likened our sample to
	high-redshifts surveys searching for clues to their nature and
	galaxy evolution.

\begin{acknowledgements}

        This paper is based on observations obtained at the
        German-Spanish Astronomical Centre, Calar Alto, Spain,
        operated by the Max-Planck Institute fur Astronomie (MPIE),
        Heidelberg, jointly with the Spanish Commission for
        Astronomy. It is also partly based on observations made with
        the Jacobus Kapteyn Telescope operated on the island of La
        Palma by the Royal Greenwich Observatory in the Spanish
        Observatorio del Roque de los Muchachos of the Instituto de
        Astrof\'{\i}sica de Canarias and the 1.52m telescope of the
        EOCA/OAN Observatory.
        
        This research has made use of the NASA/IPAC Extragalactic
        Database (NED) which is operated by the Jet Propulsion
        Laboratory, California Institute of Technology, under contract
        with the National Aeronautics and Space Administration. We
        have also use of the LEDA database,
        http://www-obs.univ-lyon1.fr.
        This research was also supported by the Spanish Programa
        Sectorial de Promoci\'on General del Conocimiento under grants
        PB96-0610 and PB96-0645. 

	We would like to thank C.E. Garc\'{\i}a-Dab\'o and S. Pascual
	for their help during part of the observing runs and
	stimulating conversations. We thank M.N. Est\'evez for
	carefully reading the manuscript and making some useful
	remarks. We are also grateful to Dr. E. Emsellem for his
	helpful comments and suggestions.

\end{acknowledgements}

\setcounter{figure}{13}
\begin{figure}
\psfig{file=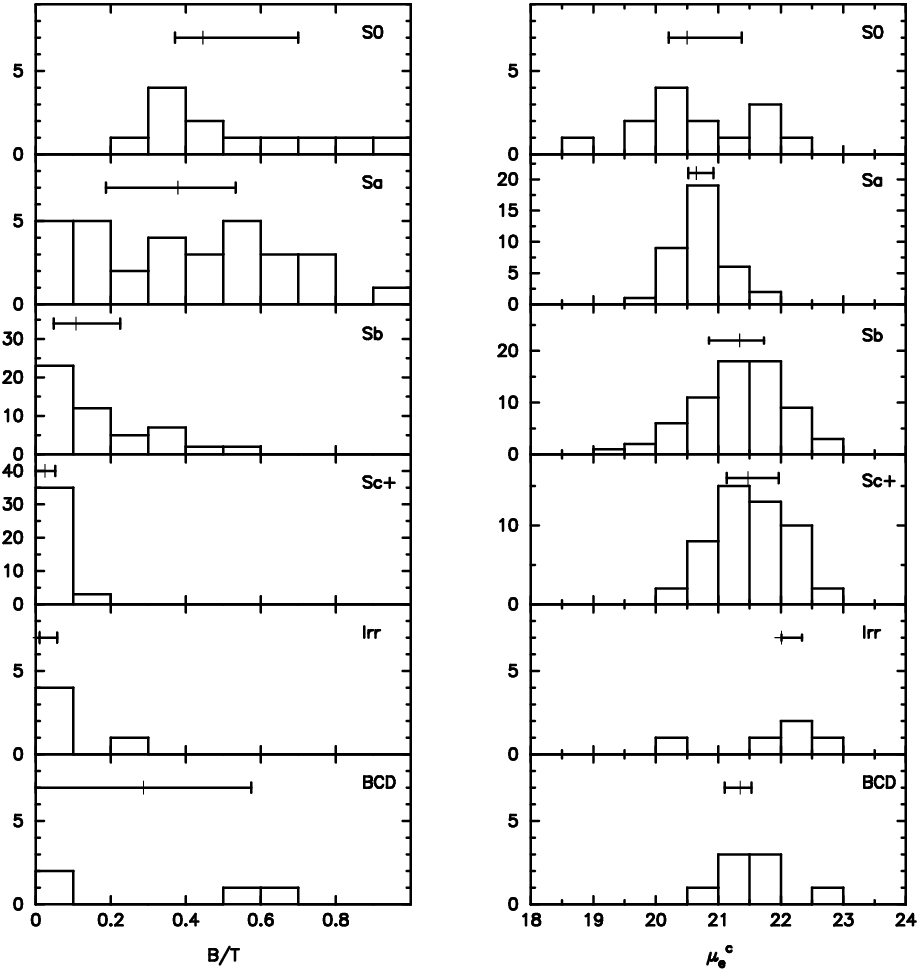,angle=0,height=7.5cm}
\psfig{file=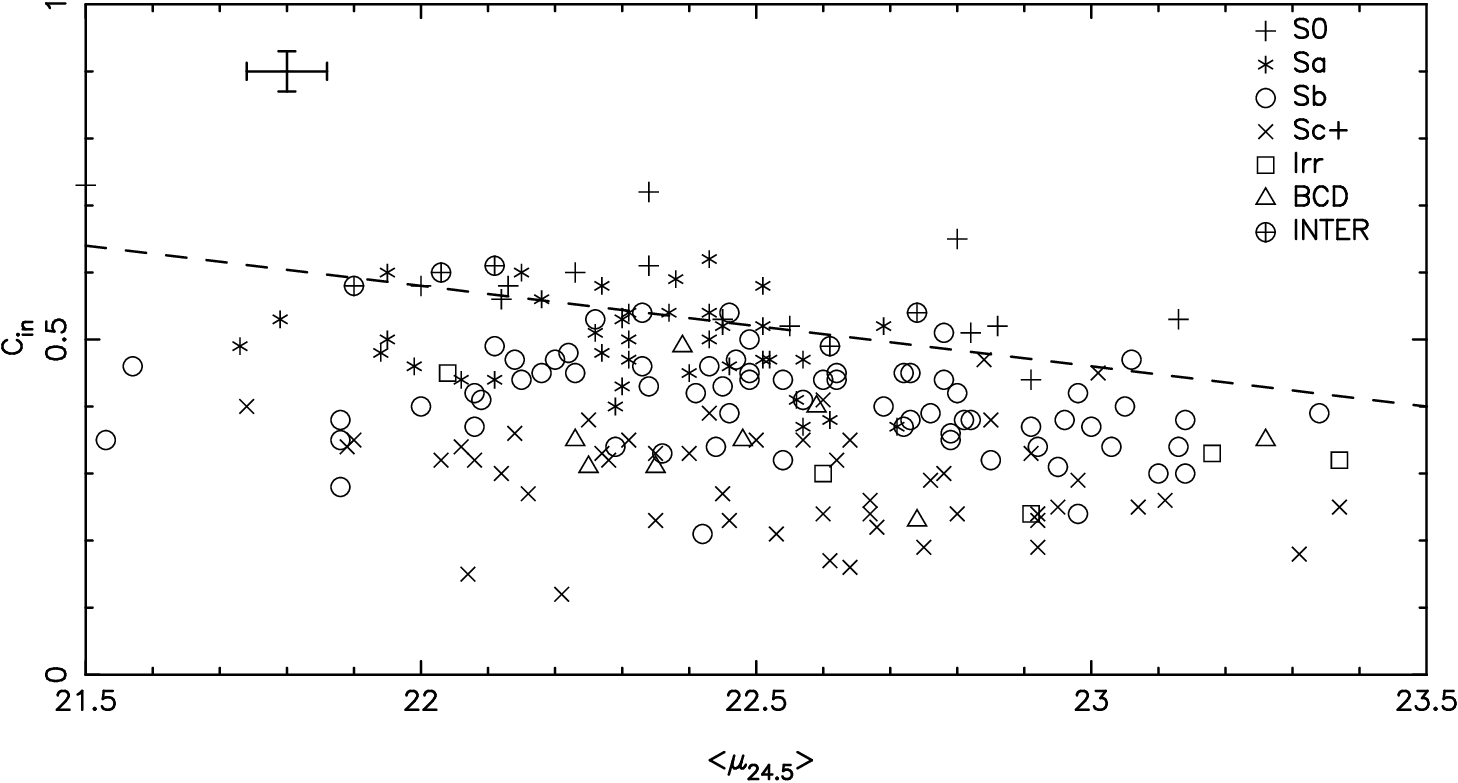,angle=-90,height=4cm}
\psfig{file=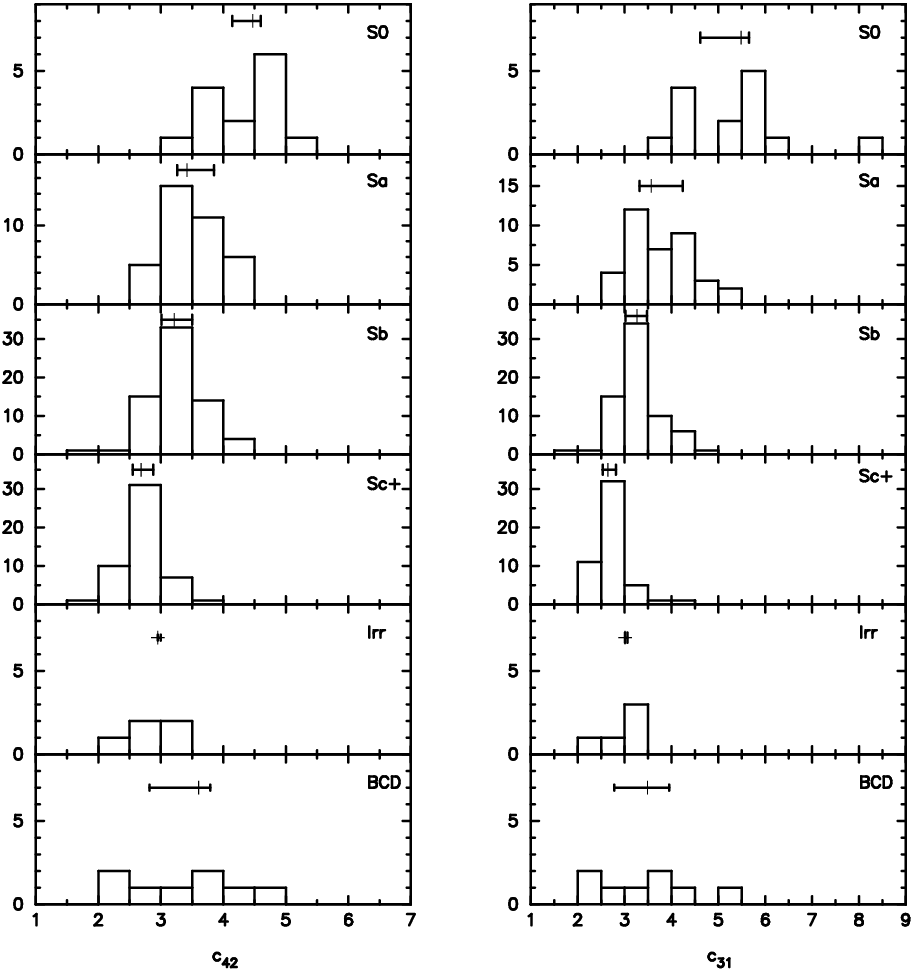,angle=0,height=7.5cm}
\caption{Plots of the 5 criteria used to morphologically classify the UCM galaxies. Top panels show on the left the distribution of the B/T ratio and on the right the mean effective surface brightness (corrected for Galactic extinction) according to the final Hubble type established for each galaxy. The middle panel shows the plot found in Doi et al. (\cite{Doi93}) of concentration index c$_{in}$ versus isophote 24.5 mean surface brightness; the dashed line represents the segregation between late-type and early-type established by Doi et al. (\cite{Doi93}). Lower panels are the histograms of the concentration indices c$_{42}$ (left) and c$_ {31}$ (right)}
\label{5crit}
\end{figure}

\end{document}